\documentclass[twocolumn,twocolappendix]{aastex701}
\usepackage{CJK}
\usepackage{fix-cm}
\usepackage{amsmath}
\usepackage{amssymb}
\usepackage{newtxtext,newtxmath}
\usepackage{array}
\usepackage{threeparttable}
\usepackage{ragged2e}
\usepackage{tabularx}
\usepackage{relsize}
\usepackage{graphicx}
\usepackage{booktabs}
\usepackage{makecell}

\begin{document}
\begin{CJK*}{UTF8}{gbsn}
\title{
  AU or pc? Inferring the distance of magnetized plasma near FRBs from propagation diagnostics}
\shorttitle{Physical scale of FRB local environments}

\author[0000-0001-5653-3787]{Wanjin Lu (陆万锦)}
\affiliation{National Astronomical Observatories, Chinese Academy of Sciences, Beijing 100101, China}
\email{wjlu@nao.cas.cn}  
\author[0000-0001-7931-0607]{Dongzi Li (李冬子)} 
\affiliation{Department of Astronomy, Tsinghua University, Beijing 100084, China}
\email[show]{dzli@tsinghua.edu.cn}

\author[0000-0002-2171-9861]{Zhen-yin Zhao (赵臻胤)} 
\affiliation{School of Astronomy and Space Science, Nanjing University, Nanjing 210093, China}
\email{zyzhao@smail.nju.edu.cn}  
\author[0000-0003-4157-7714]{Fa-yin Wang (王发印)} 
\affiliation{School of Astronomy and Space Science, Nanjing University, Nanjing 210093, China}
\affiliation{Key Laboratory of Modern Astronomy and Astrophysics (Nanjing University), Ministry of Education, Nanjing 210093, China}
\affiliation{Purple Mountain Observatory, Chinese Academy of Sciences, Nanjing 210023,  China}
\email{wangfayin@nju.edu.cn}  

\correspondingauthor{Dongzi Li}

\begin{abstract}

Fast Radio Bursts (FRBs) are highly energetic, millisecond-duration radio transients. A significant fraction of repeating FRBs are found in magneto-active environments significantly different from typical interstellar medium, offering important insights into their origins and evolutionary pathways. Possible explanations range from companion winds to young magneto-active supernvae remnants. The spatial scales of the magneto-active environment is a major distinction of different models. In this work, we present a new method to estimate the physical scale of the magneto-active region surrounding FRBs by jointly analyzing measurements of temporal scattering ($\tau_\mathrm{scat}$), depolarization ($\sigma_\mathrm{RM}$), and Faraday rotation measure (RM) variations ($\left|\Delta \mathrm{RM}/\Delta t\right|$) in repeating sources.
We systematically apply this method to all active repeaters with multiple RM measurements. 
Despite the coarse sampling and large uncertainties, the inferred distances tentatively favor SNR-scale magneto-environments for FRB 20190303A, FRB 20190417A, and FRB 20190520B, while still allowing binary-scale structures for FRB 20180916B and FRB 20201124A under plausible assumptions.
Better sampling of propagation effects, together with future advances in simultaneous wideband measurements of multiple effects with CHORD and the DSA, has the potential to systematic discrimination among the origins of FRB magneto-environments and constrain progenitor evolution.


\end{abstract}



\section{Introduction}
Fast radio bursts (FRBs) are energetic millisecond-duration radio transients that exhibit a wide range of observational properties. Since the discovery of the first repeating FRB, more than 800 sources have been detected \citep[see the statistics in \href{https://blinkverse.zero2x.org/overview}{BlinkVerse};][]{Blinkverse}, only a small fraction of which are associated with persistent radio counterparts. A variety of source models have been proposed to explain the diverse phenomenology of FRBs \citep{SGR2014, SLSN2017, KumarLu2017}, yet a unified physical picture remains under debate. One major challenge is that the observed signals are shaped not only by the emission mechanism itself, but also by propagation through magneto-ionic media along the line of sight, from the Galactic interstellar medium to the host galaxy and the immediate source environment.

The detection of the Galactic FRB-like event FRB 20200428, together with its associated X-ray burst from the magnetar SGR J1935+2154, has confirmed that magnetars can produce FRB-like radio bursts \citep{frb20a,frb20b,xray20,xray21a,xray21b,xray21c}. This source is also spatially associated with the Galactic supernova remnant (SNR) G57.2+0.8 \citep{zhou2020}, illustrating that compact objects capable of producing FRB-like emission may reside in complex and highly magnetized environments. Such environments can imprint themselves on the observed radio pulses through propagation effects \citep{Rickett1990}, thereby providing a valuable probe of the local plasma conditions around FRB sources.

One such effect is depolarization. The linearly polarized emission of many repeating FRBs shows frequency-dependent evolution \citep{feng22}, with the polarization fraction often decreasing toward lower frequencies. 
This behavior can be described by external Faraday dispersion,
$f_{\rm depol}=1-\exp(-2\sigma_{\rm RM}^2\lambda^4),$
where $\sigma_{\rm RM}$ characterizes stochastic fluctuations in the rotation measure (RM) across the emitting or scattering region \citep{burn1966}. The RM is defined as
\begin{equation}
\mathrm{RM} = \frac{e^3}{2\pi m^2_\mathrm{e}c^4}\int^d_0 n_\mathrm{e}B_\parallel dl,
\end{equation}
and traces the electron-density-weighted magnetic field along the line of sight. In this context, depolarization can be interpreted as the result of multi-path propagation through a turbulent, inhomogeneous, magnetized plasma screen surrounding the FRB source \citep{Beniamini_depol_2022,yang22}. Recent spectral studies of apparently non-repeating FRBs have also revealed significant depolarization, suggesting that such propagation effects may be common across the FRB population, regardless of repeatability \citep{askap_onoff_depol_2024,askap_onoff_depol_2025}.

Another key propagation effect is temporal scattering, which appears as an exponential broadening of the burst profile due to multipath propagation through turbulent plasma \citep{cordes_2016}. This effect is particularly prominent at low radio frequencies \citep{180916_lofar_scatter_2021,niu22}. Together, RM variations, depolarization, and scattering provide complementary diagnostics of the magneto-ionic structure of FRB environments. When monitored over long timescales, these observables can reveal the evolution, geometry, and physical scale of the plasma structures local to FRB sources.

As observational evidence accumulates, environmental interpretations of FRBs have begun to converge toward a few broad classes. Detailed studies of active repeating FRBs have revealed turbulent and dynamically evolving magneto-ionic environments \citep{michilli18,xu22,annathomas22,CHIMEdepol2023}. The secular and approximately monotonic RM evolution observed in repeaters such as FRB 20121102A and FRB 20180916B has been interpreted as evidence for the dilution of a local environment, broadly consistent with an expanding SNR scenario \citep{snr_121102_2017,SLSN2017,yang_SNR_2017,snr_121102_2021,yang_SNR_2023,180916_2023,180916_uGMRT_2024}. 
In contrast, the dramatic RM variations and reversals observed in FRB 20190520B and FRB 20201124A point to more dynamic environments, potentially involving compact objects interacting with local plasma structures such as stellar winds or flares in binary systems \citep{wangFY22,annathomas22}. 
The periodic, frequency-dependent activity window of FRB 20180916B has also motivated binary-related interpretations \citep{Periodicity_2020,180916_LOFAR_2024_dis,180916_uGMRT_2024}.
A key distinction among these environmental scenarios is their characteristic physical scale.

\begin{figure*}[ht]
\centering
\plotone{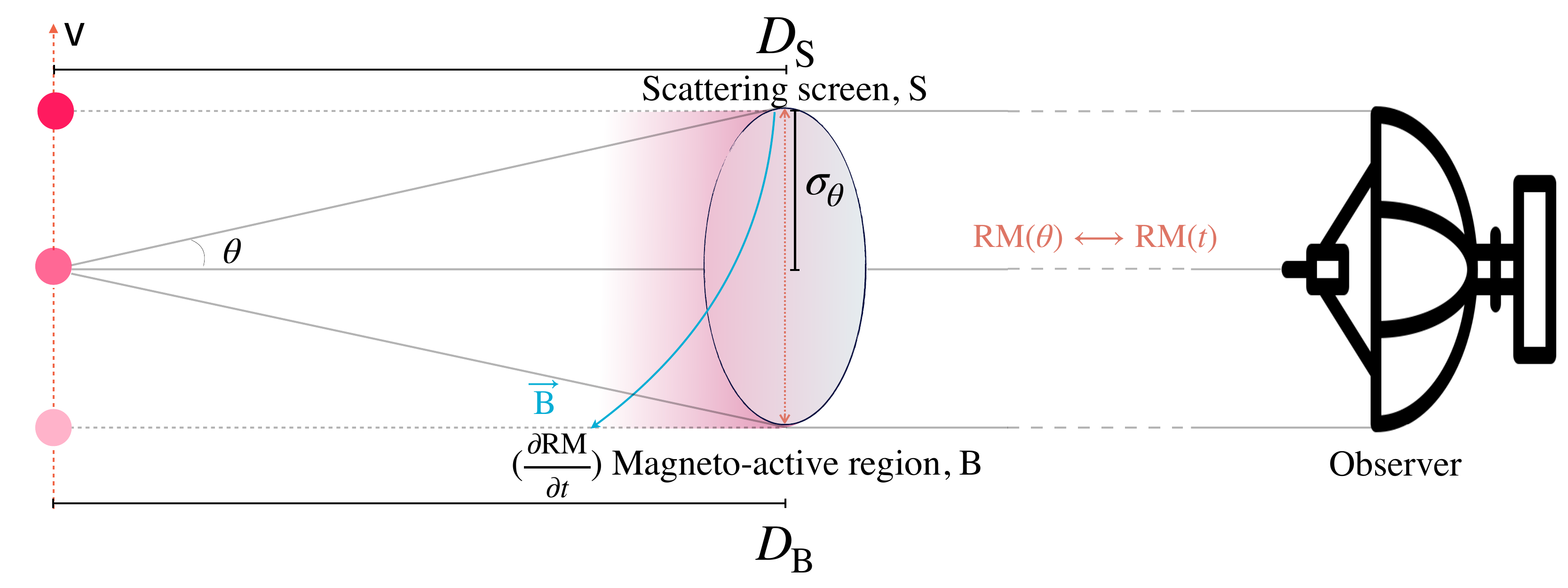}
\caption{A top-view sketch (not to scale) of the foreground screen and the moving FRB source.  
The proper motion of the FRB source is indicated as dashed arrows with a transversal velosity $\mathrm{v}$. 
Observed burst sightlines are parallel as the vicinity of the FRB source is negligible compared to the cosmological distance.}
\label{fig1}
\end{figure*}

In this paper, we provide an independent constraint on the physical scale of FRB environment with simultaneous measurements of multiple propagation effects. 
In Section~\ref{sec:Geometry}, the physical scenario of multi-path propagation effects in-situ is clarified. 
In Section~\ref{sec:application}, We apply the propagation geometry to several repeating FRB sources and estimate their physical scale.
In Section~\ref{sec:discussion}, we discuss the agreement and discrepancy between the environmental models and observations.
The results are concluded in Section~\ref{sec:conclusion}.

\section{Multi-Propagation Diagnostics of the Circumburst Magneto-Active Region}
\label{sec:Geometry}
A subset of repeaters are located in magneto-active environments with strong local scattering. 
Multiple propagation effects can be combined to probe the physical scale of the circum-source medium. 
As shown in Fig.~\ref{fig1}, a background radio source probes the surrounding magnetic structures through two complementary observables. 
First, time-dependent RM variations trace changes in the magnetic field along different lines of sight (LOS) as the source moves. 
Second, depolarization reflects RM differences among the multiple LOS associated with scattered images. 
Combining the two measurements, we give an estimate of the angular scale of the magneto-environment. 
And then the measured scattering timescale or scintillation bandwidth constrains the geometry of these scattered images and convert the angular scale estimates to physical distance estimate. 
Following this logic, we provide  detailed calculations to combine constraints from RM variability, depolarization, and scattering measurements to infer the distance between the source and the RM screen.

The schematic representation of the circumburst geometry is illustrated in Figure~\ref{fig1}. A magneto-active region, B, is located at a distance $D_\mathrm{B}$ from the FRB source and governs the RM variation. 

The FRB source (red filled circles) moves with a velocity $v$ relative to the plasma in B. The observer measures a change in RM over time as the LOS (grey dashed lines) traverses the inhomogeneous medium and samples changes in electron density $n_e$ or magnetic field $B$ within region B: 
\begin{align}
	\Delta {\rm RM} = 
	\frac{\partial \rm RM}{\partial \theta} v_\theta \Delta T + \sigma_{\rm rand}=   \frac{\partial \rm RM}{\partial \theta}\frac{\rm v}{D_{\rm B}}\Delta T + \sigma_{\rm rand}
    \label{eq:dRM}
\end{align}
where the first term represents the large-scale structure and the second term represents random fluctuations. 

At a given time, the FRB emission is scattered over an angular scale $\sigma_\theta$ (grey solid lines), producing depolarization due to RM variation across the region of B illuminated by the scattered rays. The observed linear polarization $L_{\rm obs}$ is thus reduced relative to the intrinsic linear polarization $L$ (see derivation in \cite{burn1966} and Appendix~\ref{sec:uncertainty}): 
\begin{align}
	L_{\rm obs}&=L e^{-2\sigma_{\rm RM}^2\lambda^4} \\
	&\sigma_{\rm RM}^2 = 2\sigma_{\rm \theta}^2 \left(\frac{\partial \rm RM}{\partial \theta}\right)^2 +   \sigma_{\rm rand}^2
	\label{eq:depol}
\end{align}
The factor 2 in the first term of $\sigma_{\rm RM}^2$ accounts for two-dimensional scattering by the screen (see Appendix~\ref{sec:derivation}). The derivative ${\partial \rm RM}/{\partial \theta}$ is defined along the direction of the relative velocity and therefore represents, on average, 1 / $\sqrt{2}$) of the RM variation on the two-dimensional screen.

The multi-path propagation also introduces a time delay $\tau_{\rm scat}$ between the scattered image, which is determined by the angular scale of the scattering $\sigma_\theta$ as well as the distance of the scattering screen $D_{\rm S}$: 
\begin{align}
	\tau_\mathrm{scat} = \frac{D_{\rm S}\sigma^2_\theta}{2c}
    \label{eq:tau}
\end{align}
This time delay could be measured either by modeling the scattering tail of the FRB or the scintillation bandwidth $\Delta \nu_s=2\pi/\tau_{\rm scat}$. 
Notice, the effect of scattering here is providing the angular separation of image $\sigma_\theta$ and hence the multi-path propagation. The scattering site does not necessarily corresponds to the magneto-active region B.
The separation of images are usually high frequency-dependent with $\sigma_\theta\propto \lambda^4$, therefore, it is critical to scale the $\tau_{\rm scat}(\lambda)$ to the wavelength where the depolarization is measured.



Combining Equation~\ref{eq:dRM},\ref{eq:depol},\ref{eq:tau} about the three observables, we can constrain for the distance between the source and the magneto-active region B:
\begin{align}
    D_\mathrm{B}^2=\frac{1}{4c}(\sigma_{\rm RM}^2-\sigma_{\rm rand}^2)\frac{\Delta T^2}{\Delta{\rm RM}^2+\sigma_{\rm rand}^2}v^2 \frac{D_\mathrm{S}}{\tau_{\rm scat}}
    \label{eq:DBgeneral}
\end{align}
 
Since RM is an integrated effect, we assume that the large-scale variation contribute to a significant fraction of the $\sigma_{\rm RM}$ \citep[$\sigma_\mathrm{rand}$ is negligible; also theoretically justified in ][]{yang22}.
Equation~\ref{eq:DBgeneral} simplifies to: 
\begin{align}
    D_\mathrm{B} &\sim  0.45\ \rm{pc} \left(\frac{\sigma_{\rm RM}}{10\ \rm{rad/m^2}}\right)  \left(\frac{\tau_\mathrm{scat}}{\rm 1\ ms}\right)^{-\frac{1}{2}} \left|\frac{\Delta \rm RM/\Delta t}{10\rm\ rad/m^2/day}\right|^{-1}\nonumber \\
    &\left(\frac{\rm v}{100\ \rm km/s}\right) \sqrt{\frac{D_\mathrm{S}}{100\ \rm pc}},
    \label{eq:DB}
\end{align}
Therefore, for moderate scattering in the host galaxy ISM with $D_\mathrm{S}\sim 100-1000$ pc, and moderate observed RM variation (i.e. changes no more than few hundred rad/m$^2$), the magneto-active region has to be in the physical distance $D_\mathrm{B}$ of pc scale to explain the observed depolarization. 

The potential supernovae and stellar wind that contribute to the large RM variation can also introduce local scattering. In this case the scattering screen S would locate at similar distance with the magneto-acitve region B with $D_\mathrm{S}\sim D_\mathrm{B}$ \citep{Beniamini_depol_2022,yang22}. 
In this case, the source-to-mageneto-active screen distance $D_\mathrm{B}$ can be expressed as:

\begin{align}
    D_\mathrm{B} &\sim  415\ \rm{AU} \left(\frac{\sigma_{\rm RM}}{10\ \rm{rad/m^2}}\right)^2  \left(\frac{\tau_\mathrm{scat}}{\rm 1\ ms}\right)^{-1} \left|\frac{\Delta \rm RM/\Delta t}{10\rm\ rad/m^2/day}\right|^{-2}\nonumber \\
    &\times \left(\frac{D_\mathrm{S}}{D_\mathrm{B}}\right) \left(\frac{\rm v}{100\ \rm km/s}\right)^{2}.
    \label{eq:dist}
\end{align}
Identifying the origin of the scattering site is an important task to constrain the spatial scale of the magneto-active region B in the future.

\section{The physical scales for different models}
With the above diagnostics established, we now derive the expected source-to-screen distances for two representative environmental models. 
In binary systems, the orbital motion and the stellar interactions including flares, stellar wind and the disk may significantly shape the observational properties of FRBs \citep{Periodicity_2020,RMmodel23,RMflare2025}.
The source-to-screen distance is set by the orbital separation. 
Kepler's third law gives

\begin{equation}
    D_\mathrm{B}\sim a = 0.9~\mathrm{AU}\left(\frac{M_{\mathrm{tot}}}{10~M_\odot}\right)^{1/3} \left(\frac{P}{100~\mathrm{d}}\right)^{2/3} \\
    \label{eq:binary}
\end{equation}

The variation timescale $\Delta t$ must satisfy $\Delta t \ll P_\mathrm{orb}$ otherwise the periodic evolution of RM should be detected.
The upper limit of variation timescale could be estimated as the longest orbital period observed in Galactic binaries \citep{Ho2017}. 

Moreover, the contribution of RM would decrease very fast as the binary separation a increases:
\begin{align}
    \rm{RM} \approx &a \frac{\dot{M}}{4\pi a^2 m_p} B_s \left(\frac{R_*}{R_{\rm mag}}\right)^3\left(\frac{R_{\rm mag}}{a}\right)^2\\
    \sim & 10~\mathrm{rad}/\mathrm{m}^2~\frac{\dot{M}}{10^{-7}M_\odot/{\rm yr}} \left(\frac{a}{100{\rm AU}}\right)^{-3}\nonumber \\ 
    &\times \frac{B_s}{\rm kG} \left(\frac{R_*}{10 R_\odot}\right)^3 \left(\frac{R_\mathrm{mag}}{200 R_\odot}\right)^{-1}\nonumber
\end{align}

where $\dot{M}$ is the mass loss rate of the companion, $m_p$ is the proton mass, $B_s$ is the surface magnetic field strength of the companion, $R_*$ is the radius of the companion, $R_\mathrm{mag}$ is the radius of the companion magnetosphere.  

It is difficult to introduce large RM changes when the orbital separation is orders of magnitude larger than 100 AU, unless the system is in extremely eccentric orbit. 
However, large RM change can only be observed near periastron if the system is in highly eccentric orbit, introducing periodic spikes of RM, which is not observed yet. 
Therefore, we use 100 AU as the upper limit of characteristic physical scale for the binary model. When the binary separation $a\ll$ AU, the FRB emission will introduce strong wave effects on the media making the propagation deviating from the standard behavior \citep{2020MNRAS.496.3308L}. 
Therefore, we use 1-100 AU as the characteristic physical scale for the binary model.

Supernovae explosions eject stellar materials, forming a shell structure expanding freely at velocity $V_{\mathrm{sh}} \simeq\left(2 E_0 / M_{\mathrm{ej}}\right)^{1 / 2} \simeq 10^4 \mathrm{~km} \mathrm{~s}^{-1}\left(\frac{E_0}{10^{51} ~\mathrm{erg}}\right)^{1 /2}\left(\frac{M_{\mathrm{ej}}}{1~M_{\odot}}\right)^{-1 /2}$, where $E_0$ and $M_{\mathrm{ej}}$ is the explosion energy and the ejecta mass, respectively. 
When the mass of swept ambient medium is comparable to the ejecta mass $M_{\mathrm{sw}}=4\pi\rho_0(V_{\mathrm{sh}}t)^3/3\sim M_{\mathrm{ej}}$, the evolution of the SNR enters the Sedov-Taylor phase. The duration of the free expansion phase is
\begin{equation}
\begin{aligned}
t_{\mathrm{FE}}= & \left(\frac{3 M_{\mathrm{ej}}}{4 \pi \rho_0 v_0^3}\right)^{\frac{1}{3}} \simeq 440~ \mathrm{yr}\left(\frac{n_0}{0.1 \mathrm{~cm}^{-3}}\right)^{-\frac{1}{3}} \\
& \times\left(\frac{E_0}{10^{51} \mathrm{erg}}\right)^{-\frac{1}{2}}\left(\frac{M_{\mathrm{ej}}}{1~\mathrm{M}_{\odot}}\right)^{\frac{5}{6}},
\end{aligned}
\end{equation}
where $n_0$ is the electron density of the circumstellar medium.

\begin{figure*}
\plotone{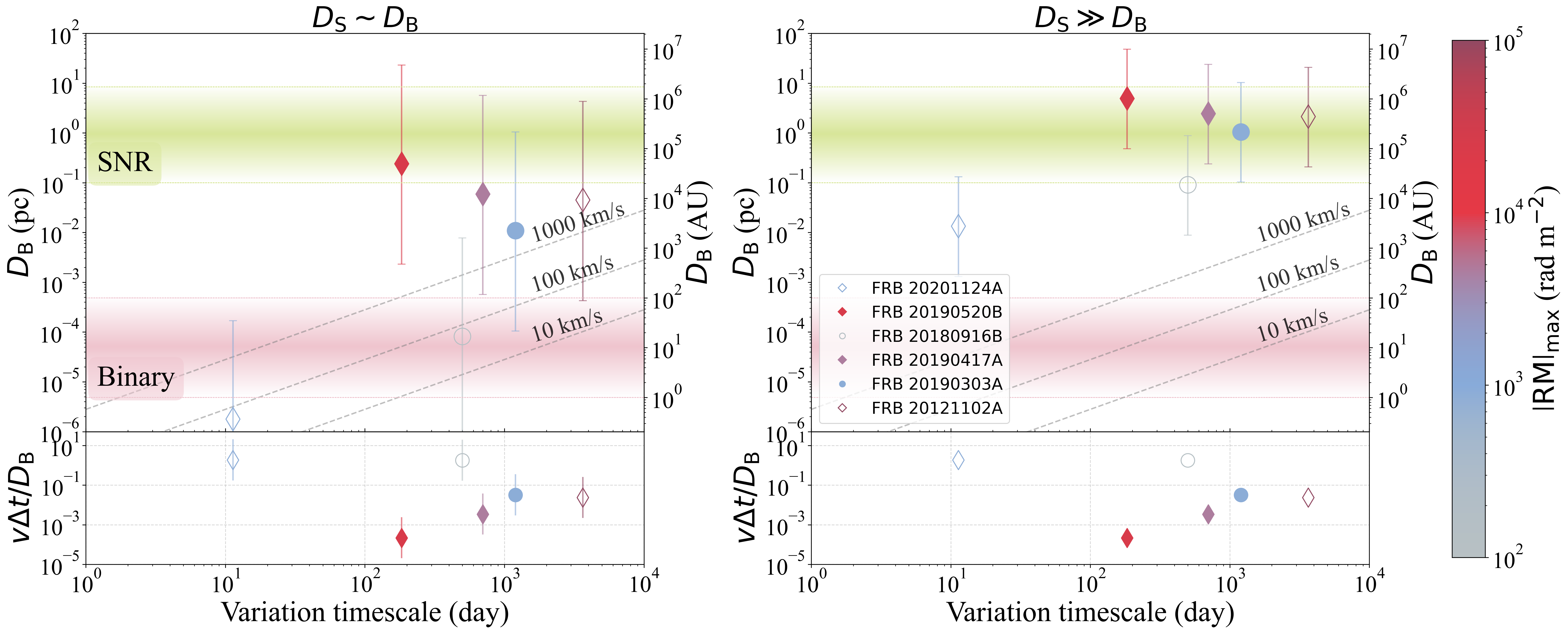}
\caption{
    The Estimate of source-to-screen distances of several repeating FRBs assuming co-location of scattering and RM variation (left) and scattering at 100\,pc (right). 
    Data points represent the distances estimated from Equation~\ref{eq:dist}, centering at a velocity of 100 km/s, with the line segment indicating the velocity spanning from 10 (lower) to 1000 (upper) km/s, respectively. 
    The diamond data points represent FRB sources that are associated with a confirmed persistent radio counterpart while the circles do not.
    The data points for FRB 20121102A, FRB 20201124A, and FRB 20180916A are shown as hollow symbols because their circum-source scattering is unknown. The observed scattering is either absent or close to the Milky Way expectation, hence the inferred distances are likely lower limits( see Appendix~\ref{sec:obsview})
    The colorbar scales logarithmically with the maximum absolute value of RM of FRB sources.
    Shaded regions are preferred parameter space for different FRB environmental models.
    The gray dashed lines represent transverse spatial scales $x_\mathrm{spatial} = v\times \Delta t$ with varying transverse velocity from 10 to 1000 km/s. 
    }
\label{fig:DVplot}
\end{figure*}

The persistent radio source associated with FRBs may originate from a pulsar wind nebula powered by a new-born magnetar \citep{Murase2016,Margalit2018,Zhao2021b}.

The spin-down or magnetic energy injection from the central compact object governs the coupled evolution of the nebula and SNR, accelerating the ejecta to speeds $\sim$ 0.1c \citep{Murase2016,Zhao2021b}. The SNR will be optically thick for $t \sim 1-10$ yr due to the free-free absorption of SN ejecta \citep{Metzger2017,Zhao2021b}.
Giving the observed maximum FRB activity duration to be $t \sim$ 10 years \citep{121102_decadal}, the source-to-screen distance must satisfy
\begin{equation}
    D_\mathrm{B}\sim R_{\mathrm{ej}}\sim R_{\mathrm{MWN}} =V_{\mathrm{ej}}t\simeq 0.3~\mathrm{pc}\left(\frac{V_{\mathrm{ej}}}{0.1~c}\right)\left(\frac{t}{10 \mathrm{~yr}}\right).
\end{equation}
The evolution of DM and RM is also dominated by the co-evolution of the MWN and the SNR. For young SNR, the evolution of RM must take into account the activity of the central engine. And it is necessary to consider multiple media contributions, such as the progenitors's ejecta, wind nebula and the SNe shell. At this point, the assumption of this paper that the variation of RM is caused by the relative motion between the source and the medium is no longer applicable.

We are more interested in the old SNR of the Sedov-Taylor phase.
The shell radius and the velocity are given by
\begin{align}
    R_{\mathrm{sh}} & =1.15\left(\frac{E t^2}{\rho_0}\right)^{1 / 5}\leq 8.4\mathrm{~pc}\nonumber\\
    & \times \left(\frac{n_0}{0.1 \mathrm{~cm}^{-3}}\right)^{-1 / 5}\left(\frac{E_0}{10^{51} \mathrm{erg}}\right)^{1 / 5}\left(\frac{t}{1000 \mathrm{~yr}}\right)^{2 / 5}\\
    V_{\mathrm{sh}} & =\frac{2}{5} \frac{R_{\mathrm{sh}}}{t} =3300\mathrm{~km} \mathrm{~s}^{-1}\nonumber \\
    & \times \left(\frac{n_0}{0.1 \mathrm{~cm}^{-3}}\right)^{-1/5}\left(\frac{E_0}{10^{51} \mathrm{erg}}\right)^{1 / 5}\left(\frac{t}{1000 \mathrm{~yr}}\right)^{-3/ 5}.
\label{eq:snr}
\end{align}

Therefore, we use 0.1-10 pc as the characteristic physical scale for the supernovae remnant. 

\section{Applications on repeating FRB sources}\label{sec:application}

We now apply the multi-propagation diagnostic framework developed in Section~\ref{sec:Geometry} to a sample of active repeating FRBs with available measurements of RM variation, depolarization and scattering. 
In this section, we present the parameter selection strategy and the estimated source-to-source distances.
As a secondary consistency check, we also examines a log-linear scaling relation that is naturally predicted under single-class origin. 

\subsection{Constraining the physical scale of FRB local environments}

{\setcitestyle{numbers}
\begin{table*}
    \centering
    \caption{Observational properties including $\tau_\mathrm{scat}$ and $\sigma_\mathrm{RM}$ of some FRB sources}
    
    \begin{tabularx}{0.8\textwidth}{@{}lccccl@{}}
    \hline
    \hline
    Object & RM & $\left|\Delta \mathrm{RM}/\Delta t\right|^\dagger$ & $\tau^\mathrm{obs}_\mathrm{scat}$  &$\sigma_\mathrm{RM}$  & reference \\
           & $\rm rad/m^2$ & $\rm rad/m^2/day $ & (ms) &  $\rm rad/m^2$ & \\
    \hline
    {FRB 20190520B} &  -23997 -- 12523 & 7.5 $\times 10^1$ & $10$@1.3 GHz & $166.4\ \pm\ 27.5$ & \cite{feng22,annathomas22,niu22} \\
    {FRB 20201124A} &  -889.46 -- -365.12 & 2.6 $\times 10^1$ & $11$@600 MHz & $5.1$ & \cite{xu22,lu23} \\
    {FRB 20121102A} &  66949 -- 102708 & 2.8 $\times 10^1$ & $<0.43$@600 MHz  & $30.9\ \pm\ 0.4$  & \cite{feng22,CHIMEcat,121102_decadal}\\
    {FRB 20180916B} &  -120 -- -50$^*$ & 1.4 $\times 10^{-1}$  & $51$@110 MHz & $0.17\ \pm\ 0.01$ & \cite{feng22,180916_2023,180916_uGMRT_2024,180916_lofar_scatter_2021} \\
    {FRB 20190303A} &  -700 -- -500 & 1.2 & $5.5\ - 27$@600 MHz  & $3.6\ \pm\ 0.1$  & \cite{feng22,CHIMEdepol2023,CHIMEcat,KSPpol2025} \\
    {FRB 20190417A} &  4429 -- 4614 & $\sim 1$ & $17\ - 26$@600 MHz & $5.19\ \pm\ 0.09$  & \cite{feng22,CHIMEdepol2023,CHIMEcat,KSPpol2025} \\
    \hline
    \end{tabularx}
    \tablecomments{$\mathrm{^\dagger}$ The differential RM changing rates are obtained from selected burst samples (see in Section~\ref{sec:obsview}) and adopted from corresponding references.\\
    $\mathrm{^\P}$ The scattering timescales at their observing frequency are marked.\\
    $\mathrm{^{*}}$ RM range is artificially estimated from corresponding references and \citet{121102_decadal} on the decadal evolution of FRB 20121102A.
    }.
    \label{table:info}
\end{table*}
}

The source-to-screen distance $D_\mathrm{S}$ can be estimated from Equation~\ref{eq:dist} for each source in Table~\ref{table:info}, adopting a fiducial transverse velocity of 100 km/s flexible from 10-1000 km/s. 
The RM variation rate $\left|\frac{\partial \mathrm{RM}}{\partial t}\right|$ is approximated by $ \left|\frac{\Delta \rm RM}{\Delta t}\right|$ over a characteristic timescale $\Delta t$, defined as either the first break in the structure function (SF) of the RM time series, or determined by manual inspection when the data cadence is insufficient for SF analysis. 
For sources with secular RM evolution, $\Delta t$ corresponds to the full monitoring baseline; for sources exhibiting dynamic multi-stage variations, representative sub-intervals are adopted. 
The scattering timescale is assumed stable over the RM variation baseline and is extrapolated to the critical frequency $\nu_{50}$ where linear polarization is reduced to 50\%, using a power-law index  $\alpha = -4$. 

For physical interpretation, we also calculate $x_\mathrm{spatial}=v\Delta t$, the projected distance traversed by the source during its RM variation timescale. 
As shown in the left panel of Figure~\ref{fig:DVplot}, the ratio $x_\mathrm{spatial}/D_\mathrm{S}$ provides a useful diagnostic: 
in a binary system, the ratio approximates the orbital eccentricity, since the sources sweeps across a significant fraction of the orbital separation during one variation cycle; 
in an SNR, by contrast, the natal kick velocity of the central compact object is negligible compared to the shell expansion, yielding $x_\mathrm{spatial}/D_\mathrm{S} \ll 1$. 

The right panel of Figure~\ref{fig:DVplot} considers another scenario in which the scattering screen is spatially decoupled from the magneto-active region. 
Assuming that the observed scattering originates from a distant host-galaxy structure \citep[e.g. $D_\mathrm{S} \sim 100\ \rm pc$; ][]{ocker22_DM}, $D_\mathrm{B}$ from the source to magneto-active region is estimated from Equation~\ref{eq:DB}. It is possible that the assumed $D_\mathrm{S}$ differs by an order of magnitude from the actual value. However, it would only influence the final result by a factor of few as $D_{\rm B}\propto \sqrt{D_\mathrm{S}}$, and therefore, only have minor impact on our interpretation. 

The resulting $D_\mathrm{B}$ estimates reveal a clear differentiation among the six sources. 
FRB 20121102A, FRB 20190303A, FRB 20190417A, and FRB 20190520B lie in the SNR regime. 
FRB 20201124A and FRB 20180916B fall within the binary parameter space, consistent with RM variations observed on timescales from days to years and with their episodic activity. 
However, their inferred screen distances would instead favor an SNR-like interpretation if the scattering screen and the magneto-active region are spatially decoupled.


\subsection{Detailed observational properties of each FRB source}
We describe below the observational inputs adopted for each source. 

{\it FRB 20201124A} and {\it FRB 20190520B}: 
Their observed RM variations occur in distinct phases with differing rates and durations, yet since all values vary by less than an order of magnitude, we employ representative values for the estimates here. 
The long-term observations of FRB 20190520B were split into three stages to obtain the RM differential changing rate (see details in Appendix~\ref{sec:obsview}).
We adopted the RM variation rate and the depolarization value in the first stage of FRB 20190520B as shown in Figure~\ref{fig: 190520B} when there is measurement of scattering time at similar time.
For FRB 20201124A, the RM changing rate was chosen as 25 rad/m$^2$/day similarly. Full RM variation curve is plotted as Figure~\ref{fig: 201124A}.
Besides, its daily changing rate of RM was observed to show significant variation from 0.1 to over 100 rad/m$^2$/day.    
We plotted the intra-burst RM difference against the waiting time to decouple the intrinsic RM variation from the measurement uncertainty but failed, since that the distribution of RM variations are dominated by the bi-peak waiting time clusters and therefore widely dispersed.
Nevertheless, the source-to-screen distance estimates remain robust at the order-of-magnitude level despite the limited cadence and rapid RM variability from these two FRBs.
The depolarization of FRB 20201124A was chosen as the median value (5.1 rad/m$^2$) of the daily depolarization from \cite{lu23}. 
The variation timescales of these two FRB sources are identified as the first peak in the structure functions of their observed RM variation curves as shown in Figure~\ref{fig: SF}.
Their scattering timescales were measured by CHIME and FAST, respectively, and maintained stable throughout their multi-epoch observations.

{\it FRB 20180916B}: 
The corresponding RM changing rate turns out to be 3.35 rad/m$^2$/day by linear fitting, combining all the FRB samples with RM measurements from MJD = 59250 to 59850. 
The time span of the RM decrease reported in \cite{180916_uGMRT_2024} is adopted as its variation timescale. 
The scattering timescale measured by LOw Frequency ARray (LOFAR) at 150 MHz maintained constant throughout the observations \citep{180916_lofar_scatter_2021,180916_LOFAR_2024_dis}.

{\it FRB 20190303A} and {\it FRB 20190417A}: 
We combined the RM observations of CHIME and FAST and fitted the variation curve in several sub-stages using linear function in Figure~\ref{fig: 190303A} and Figure~\ref{fig: 190417A}, respectively \citep{CHIMEdepol2023,KSPpol2025}.
For FRB 20190303A, the higher RM changing rate (the later stage) was adopted.
The variation timescale of FRB 20190303A is estimated to be 700 days, which is the time span of all the RM observations.
Given that the RM variation of FRB 20190417A follows a seeming raise-and-fall trend within the same orders of magnitude, the linear fitted RM differing rate and the time span across all the observations were adopted. 

{\it FRB 20121102A}: 
Recent L-band observations show that the RM of FRB 20121102A has decreased by$\sim 70 \%$ \citep{121102_DoL_Lband_atel,decadal_2025}.
Combined with decadal observations, the RM of FRB 20121102A decreases persistently at a rate of 28.5 rad/m$^2$/day, with a corresponding variation timescale of 10 years.

The scattering timescales of All the FRB sources are extrapolated from the observing frequency to their critical depolarization frequency as shown in Table~\ref{table:tau_freq}.

\begin{figure}
\plotone{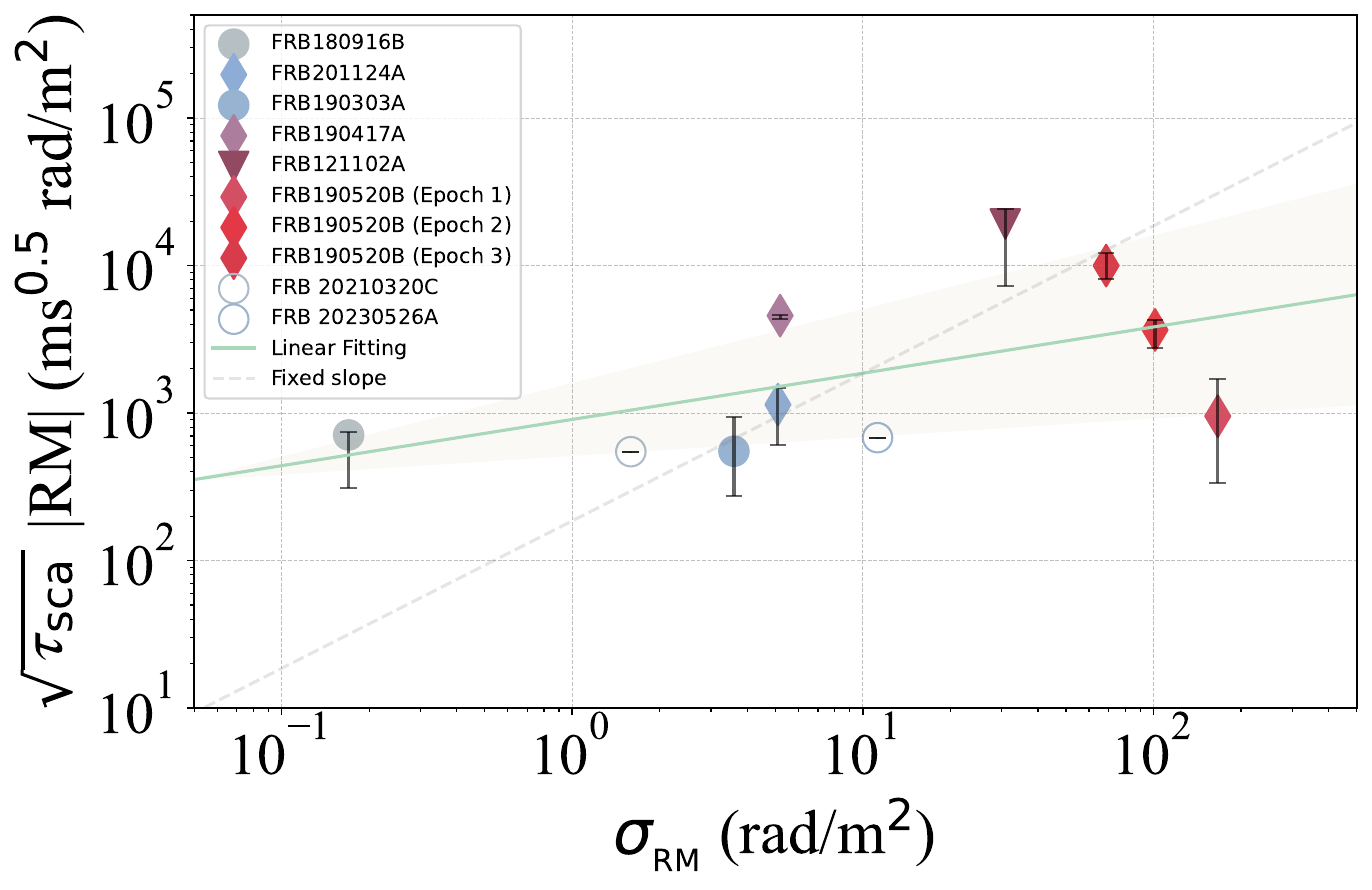}

\caption{The $\sqrt{\tau_\mathrm{scat}}$ $|$RM$|$ -- $\sigma_\mathrm{RM}$ distribution for several repeating FRBs. 
Solid data points represent FRBs associated with a persistent radio source while the hollow ones do not.
Lower triangle is applied to indicate the upper limit of the scattering timescale of FRB 20121102A \citep{CHIMEcat}.
Circles are two non-repeating FRBs detected by the Australian SKA Pathfinder (ASKAP) with depolarization measurements \citep{askap_onoff_depol_2025}. 
Colors of each data point stand for the maximum value of $|$RM$|$, sharing the same log-scale colorbar with Figure~\ref{fig:DVplot}.
FRB 20190520B (red diamonds) is split into three stages to show the tentative variation (see in Appendix~\ref{sec:obsview}). }
\label{fig:estimate}
\end{figure}

\subsection{A dynamic evolution path of FRBs}

Equation~\ref{eq:dist} could be rearranged to a log-linear relation between combined propagation observable, providing a graphical consistency check as
\begin{equation}
{\rm log}(\frac{\sqrt{\tau_\mathrm{scat}}|{\rm RM}|}{\mathrm{ms}^{\frac{1}{2}}\mathrm{rad/m}^2}) = {\rm log}(\frac{\sigma_\mathrm{RM}}{\mathrm{rad/m}^2} )+ \frac{1}{2}[{\rm log}(\frac{x^2_\mathrm{spatial}/D_\mathrm{S}}{4\ \mathrm{AU}}) +2].
\label{eq:log-linear}
\end{equation}
Figure~\ref{fig:estimate} shows the resulting $\sqrt{\tau_\mathrm{scat}}$ $|$RM$|$ -- $\sigma_\mathrm{RM}$ distribution for the selected repeating FRBs, along with two non-repeating FRBs from \cite{askap_onoff_depol_2025}. 
The data points show a weak positive trend, though the fitted slope is $0.31^{+0.19}_{-0.19}$ is shallower than the theoretically expected value of unity and the scatter is large. 
The data are broadly compatible with Equation~\ref{eq:log-linear}, which may indicate that the FRBs undergo a dynamic evolution process or host multi-class origins.

\section{Implications on the environments of FRBs}\label{sec:discussion}
The source-to-screen distances inferred in Section~\ref{sec:application} span several orders of magnitude from AU to pc, without being anchored to any prior assumption about the FRB environment. 
Rather than testing whether observations are consistent with a pre-specified model, this  diagnostic framework directly combine the multiple propagation observables to suggest the relevance to different physical scales. 
Nevertheless, $D_\mathrm{S}$ is an order-of-magnitude estimate subject to several assumptions, and the inferred values do not map cleanly onto a single environmental scenario for all sources. 
In this section, we examine the degree of consistency and tension between the inferred scales and the two leading environmental models, and discuss sources of uncertainty that limit the interpretability of the current results.

\subsection{Comparison with binary and SNR environments}
For a binary origin, RM variations are expected to be periodic, modulated by the orbital motion. 
The passage through the companion's decretion disk near the periastron should further produce periodic sign reverals in RM \citep{wangFY22}. 
However, multi-year monitoring of FRB 20201124A and FRB 20180916B have yet to detect any orbital periodicity, which may indicate an exceptionally long orbital cycle extending beyond the current observational baseline \citep{xujw2025}. 
Additionally, the RM behavior in binary system may be further complicated by non-orbital effects.
A very recent study on FRB 20220529A discovered that stellar flares can produce dramatic RM leap that may mimic or obscure the orbital signature \citep{RMflare2025}.

For the SNR scenario, the secular dilution of the surrounding medium predicts a coordinated, monotonic decline in both DM and RM \citep{Margalit2018}. 
While this behavior is observed in FRB 20121102A over decadal timescales \citep{decadal_2025}, the two SNR candidates identified in Figure~\ref{fig:DVplot}, FRB 20190303A and FRB 20190417A, show RM variations that are roughly stable in magnitude but irregular in trend. 
More critically, their observed fractional RM variability $\left | \frac{\delta \mathrm{RM}}{\mathrm{RM}} \right |$ significantly exceeds the value predicted for evolved SNRs (Equation~\ref{eq:RMfrac_SNR}), though it may be less constraining during the free-expansion phase. 
These discrepancies suggest that either these sources are at an earlier evolutionary stage than assumed, or that additional local structures contribute to the observed RM fluctuations. 

An alternative, though not mutually exclusive, interpretation invokes inhomogeneous filaments and clumps within the SNR or binary environment. 
The crossing timescale of an AU-sized clump is estimated to be several years (Equation~\ref{eq:t_clumps}), comparable to the observed RM variation timescales of these sources. 
Such transient clump passages could account for the $|$RM$|$ decrease observed in FRB 20180916B, as well as the irregular RM fluctuations in FRB 20190303A and FRB 20190417A, without requiring a global evolution of the surrounding medium. 

\subsection{Systematic uncertainties in the distance diagnostic}
In practice,  $|\frac{\Delta \rm RM}{\Delta t}|$  serves as a cadence-limited observational proxy for the intrinsic RM gradient $\left|\partial \mathrm{RM}/\partial t\right|$. 

Two additional sources of uncertainty may affect the reliability of the inferred $D_\mathrm{B}$. 
First, FRB sources in highly active environments may exhibit significant stochastic RM noise (Equation~\ref{eq:sigRM_noise}), leading to an overestimate of $\sigma_\mathrm{RM}$ and, consequently, a non-negligible overestimate of $D_\mathrm{B}$. 
Second, if depolarization is dominated by large-scale structures spatially decoupled from the scattering screen, the angular scale of the depolarization screen acquires a frequency dependence $\sigma_\theta \sim \sigma_0 (\frac{\lambda}{\lambda_0})^2$, and the resulting depolarization follows a steeper $\lambda^8$ scaling rather than $\lambda^4$.
\begin{equation}
\frac{\left| L' \right|}{\left| L \right|} = e^{-2\sigma^2_\theta(\frac{\partial \mathrm{RM}}{\partial \theta})^2\lambda^4} = e^{2(\sigma^\mathrm{LS}_\mathrm{RM})^2\lambda^{-4}_0\lambda^8}.
\label{eq:depol_LargeScale}
\end{equation}
Within a limited observing bandwidth, these two scalings are difficult to distinguish observationally. 
Disentangling the physical origin of depolarization will therefore require broadband, multi-frequency polarization measurements.

A further caveat concerns the origin of the scattering timescale $\tau_\mathrm{scat}$ entering Equation~\ref{eq:dist}. 
We have assumed that scattering originates from the same local site as the RM variations. 
However, if scattering is dominated by the Galactic ISM, $\tau_\mathrm{scat}$ from a given local circum-source environment would be overestimated, leading to an underestimation of $D_\mathrm{B}$. 
Alternatively, the scattering screen may reside in the host galaxy but at a large distance from the magneto-active region — a situation that appears relevant for FRB 20190520B, whose scattering is thought to be dominated by its host galaxy environment \citep{ocker22_DM,ocker23_scatter}, and possibly for FRB 20180916B, where low-frequency observations suggest that scattering and depolarization are spatially uncorrelated \citep{180916_LOFAR_2024_dis}. 
According to the Galactic scattering estimated from the electron density models listed in Table~\ref{table:tau_freq}, the measured scattering timescale of FRB 20180916B cannot be ruled out as being dominated or significantly contributed by scattering in the Milky Way, which would result in an underestimation of $D_\mathrm{B}$. 
The separated estimates on the $D_\mathrm{B}$ shown in the right subplot of Figure~\ref{fig:DVplot} suggest that the independent screens scenario could not be excluded.

Although FRB 20121102A consistently occupies the SNR-like region of the parameter space, this correspondence should be interpreted with caution, because its RM evolution may be driven by the rapid expansion of a local supernova remnant rather than by the proper motion of FRB source assumed in our framework.

\medskip
In summary, these tensions identified above indicate that the inferred $D_\mathrm{B}$ values resist simple classification into a clean dichotomy of FRB environmental models. 
Some sources overlap with the expected scales of binary systems, others with those of supernova remnants.
Furthermore, the varying locations of FRB 20180916B, FRB 20190520B, and FRB 20201124A suggest that their scattering screens and magneto-active regions may arise from physically distinct regions.
Rather than forcing a categorical assignment, we emphasize that the wide spread of inferred distances itself reflects a genuine diversity in the origins and evolutionary stages of repeating FRBs. 

\section{Conclusion}\label{sec:conclusion}
We have presented a new method to constrain the physical scale of magnetized FRB environments by jointly analyzing three propagation effects: RM variation, depolarization, and temporal scattering. 
Critically, this framework does not assume a specific environmental model at the outset; instead, it infers the source-to-screen distance $D_\mathrm{S}$ directly from observational quantities, providing a model-independent scale that can subsequently be compared against theoretical predictions.

Applying this method to active repeating FRBs with available long-term propagation measurements, we observe 
FRB 20190303A, FRB 20190417A and FRB 20190520B favor an SNR-scale environment, whereas a binary interpretation remains viable for FRB 20180916B and FRB 20201124A under plausible assumptions. 
Similarly, FRB 20121102A appears SNR-like in parameter space, yet its RM evolution may instead be dominated by rapid expansion of a local supernova remnant rather than by the source proper motion assumed in our framework. 

Currently, the three propagation effects are measured only coarsely in time and often not simultaneously, and local scattering remains unconstrained for three sources. These results are therefore only suggestive and carry substantial uncertainties. However, with more regular monitoring and upcoming wideband instruments, this method could become a helpful diagnostic tool.
This framework highlights several observational priorities. 
Independent constraints on scintillation screen distances, such as those enabled by scintillation analyses \citep{Scin_Kumar24,ScinArc_Nimmo25}, would provide a direct external anchor for the inferred $D_\mathrm{S}$ and offer a powerful test of whether the scattering and Faraday-active regions are physically associated. 
Equally important are long-term, broadband, and preferably contemporaneous measurements of RM, depolarization, and temporal scattering, which are needed to track the coupled evolution of these propagation effects without conflating intrinsic source variability with propagation through the local environment. 
As the sample of well-characterized repeating FRBs continues to grow, such coordinated observations will allow this diagnostic framework to map the diversity of FRB environments and to distinguish ordinary evolutionary pathways from genuinely exceptional local conditions.



\section{Acknowledgment}
The authors thank for fruitful discussions with Wenbin Lu, Robert A. Main, Yuanpei Yang, Bing Zhang.


\bibliography{ref}
\bibliographystyle{aasjournalv7}

\appendix

\section{Mathematical Derivation of Propagation Equations} \label{sec:derivation} [Here, provide all the step-by-step derivations and mathematical justifications.]
The depolarization effect can be derived from \cite{burn1966} by performing a first-order expansion of RM along the line-of-sight. 
In this context, $\frac{\partial \rm RM}{\partial \theta}$ represents the large scale RM gradient of the screen: 
\begin{align}
    L' &= \frac{1}{\sqrt{2\pi}\sigma_\theta}\int L e^{i\ 2{\rm RM}(\theta)\lambda^2}\ e^{-\frac{\theta^2}{2\sigma^2_\theta}}{\rm d}\theta\nonumber \\
    &= \frac{1}{\sqrt{2\pi}\sigma_\theta}\int L e^{i\ 2({\rm RM}_0 + \frac{\partial \rm RM}{\partial \theta}\theta)\lambda^2}\ e^{-\frac{\theta^2}{2\sigma^2_\theta}}{\rm d}\theta\nonumber \\
     &=  L\ e^{2i {\rm RM}_0\lambda^2}\ e^{-2\sigma^2_\theta \left(\frac{\partial \rm RM}{\partial \theta}\right)^2 \lambda^4},\nonumber\\
     \frac{\left| L' \right|}{\left| L \right|} &= e^{-2\sigma^2_\theta (\frac{\partial \rm RM}{\partial \theta})^2 \lambda^4} 
\end{align}
where $L'$ is the observed (depolarized) linear polarization with respect to intrinsic linear polarization $L$. 
The exponent $\sigma_\mathrm{RM} = \sigma_\theta (\frac{\partial \rm RM}{\partial \theta})$ quantifies the fluctuation of RM screen in the direction of $\theta$. 

Turbulent fluctuation introduce an additional stochastic Faraday rotation $e^{i\Delta \phi}$, uncorrelated with angular position: 
\begin{align}
    L' &= \frac{1}{2\pi\sigma_\theta\sigma_\phi}\int L e^{i\Delta_\phi} e^{-\frac{\Delta \phi^2}{2\sigma^2_\phi}} e^{-\frac{\theta^2}{2\sigma^2_\theta}}{\rm d}\theta{\rm d}\left(\Delta \phi\right)\nonumber \\
    &= \frac{1}{2\sqrt{\pi}\sigma_\phi}\int L e^{i\Delta_\phi} e^{i\ 2{\rm RM}(\theta)\lambda^2}\ e^{-\frac{\Delta \phi^2}{2\sigma^2_\phi}}{\rm d}\left(\Delta \phi\right)\nonumber \\
    &= Le^{2i\phi_0}e^{-\frac{\sigma^2_\phi}{2}},\quad \frac{\left| L' \right|}{\left| L \right|} = e^{-\frac{\sigma^2_\phi}{2}} 
\label{eq:sigRM_fluctuation}
\end{align}

Such random RM noise may bias the inferred screen distances $D_\mathrm{S}$ and $D_\mathrm{B}$, particularly when the screen is magneto-active and dynamic (see Section~\ref{sec:discussion} for further discussion). 
Combining the large-scale RM gradient with this stochastic RM noise $\sigma_\mathrm{rand}$, the total depolarization effect becomes

\begin{align}
    L' &= \scalebox{0.8}{$\frac{e^{i\phi_0}e^{i\ 2{\rm RM}_0 \lambda^2}}{2\pi \sigma_\phi\sigma_{\theta}}\iint L e^{\Delta \phi}\ e^{i \frac{\partial \rm RM}{\partial \theta}\theta\lambda^2} e^{-\frac{i\Delta \phi^2}{2\sigma^2_\phi}} e^{-\frac{\theta^2}{2\sigma_\theta^2}}{\rm d}\theta \mathrm{d}(\Delta \phi)$}\nonumber \\
    &= e^{i(\phi_0+2{\rm RM}_0\lambda^2)}Le^{-2\sigma^2_{\rm rand}\lambda^4} e^{-2\sigma^2_\theta (\frac{\partial \rm RM}{\partial \theta})^2 \lambda^4} \nonumber \\
    \frac{\left| L' \right|}{\left| L \right|} &= e^{2(\sigma^\mathrm{Rec}_\mathrm{RM})^2\lambda^4},\mathrm{where}\  \sigma^\mathrm{Rec}_{\rm RM} = \sqrt{\sigma^2_\theta (\frac{\partial \rm RM}{\partial \theta})^2 + \sigma^2_{\rm rand}} 
\label{eq:sigRM_noise}
\end{align}
where $\sigma^{\rm Rec}_{\rm RM}$ is the reconstructed RM variation, encompassing both the gradient-induced and intrinsic RM random contributions in the screen. 

Tthe RM gradient of the screen at a specific direction should be the total derivative with respect to two orthogonal directions:
\begin{equation}
    \sigma_\theta(\frac{\partial \rm RM}{\partial \theta})^2 = \sigma_x(\frac{\partial \rm RM}{\partial x})^2 + \sigma_y(\frac{\partial \rm RM}{\partial y})^2
\end{equation}
Assuming a symmetric RM distribution, the total depolarization metric could be simplified as introduced in Equation~\ref{eq:depol}. 
In practice, the proper motion of the FRB source projected on the foreground scattering screen is one-dimensionally sampled as a function of angular distance $\theta$ as it moves.
Therefore, the observed RM fluctuation is expected to be $\frac{1}{\sqrt{2}}$ smaller than 
\section{Demonstration via simulated foreground screens}

The $D_\mathrm{S}$ of these three FRBs could be also estimated from simulations based on the observed RM variation curves.
The propagation of FRB through the scattering screen should be stable within a specific time window. 
When the FRB passes through the screen, only a small fraction of the screen is illuminated and therefore dominates the propagation effects.
The idea of the illuminated region and the whole screen, is similar with the concept of seeing and the whole atmosphere in ground-based astronomy.
We performed a series of simulations (see Appendix~\ref{sec:simulation}) by Gaussian sampling the simulated screen with different screen trials.
The $D_\mathrm{S}$ could be estimated as $\frac{(vt_\mathrm{L})^2}{2c\tau_\mathrm{scat}}$, where $t_\mathrm{L}$ is the best-correlated illuminated region size.

\subsection{Simulation method}\label{sec:simulation}
To reconstruct the spatial structure of the foreground plasma screen, a well-defined RM gradient and the sampling method are required.
The RM oscillation of repeating FRB 20201124A was interpreted as a periastron passage in a magnetar/Be binary system in dozens of days \citep{wangFY22}. 
Assuming the RM gradient of the scattering screen to be continuous, we then took the fitted RM variation above to be the 1D RM template $\mathrm{RM}(t)$ as the upper panel in the left subplot of Figure~\ref{fig: screen_total} is shown. 
Either the fitted binary model or the interpolation function is acceptable since both of them provide a smooth continuous RM template for the simulation.

Firstly, we generated the 2D screen $\mathrm{RM}(x,y)$ by simply adopting the horizontal axis as the actual moving direction (i.e. $\mathrm{RM}(x,y)=\mathrm{RM}(x)=\mathrm{RM}(t)$ ) so that the hypothetical FRB source with no vertical motion would reconstruct the exact 1D RM template.
The propagation of FRB is dominated by a small region around the projected location rather than the whole foreground interstellar medium. 
As is introduced in Equation~\ref{eq:depol}, this propagation region should be Gaussian sampled.
Secondly, we applied a square Gaussian filtering kernel as the target region to sample the RM screen along the moving track. 
The mean value and Gaussian weighted standard deviation were calculated and recorded as the observed RM and it standard deviation $\sigma^\mathrm{obs}_\mathrm{RM}$, simultaneously. 
Nevertheless, the observed RM variation is a 1D time series. The $\sigma_\mathrm{RM}$ is therefore squeezed as 1D Gaussian sampled, namely reconstructed standard deviation $\sigma^\mathrm{Rec}_\mathrm{RM}$. 

As the FRB source moves, the screen distance $D_\mathrm{B}$ could be inferred with $\sigma_\mathrm{RM}$ measured and certain velocity determined according to Equation~\ref{eq:dist}.
It should be clarified that the direction of proper motion $x'$-axis may not be parallel with the direction of RM gradient $x$-axis at a inclination angle $\alpha$: 
\begin{equation}
\left\{
\begin{aligned}
    \theta'_x &=  \theta_x{\rm cos \alpha}  +  \theta_y{\rm sin \alpha}  \nonumber \\
    \theta'_y &= -\theta_x{\rm sin \alpha}  + \theta_y{\rm cos \alpha}.\nonumber
\end{aligned}
\right.
\end{equation}
Therefore, the RM gradient on the actual x-axis  $\frac{\partial \rm RM}{\partial \theta_x} = \frac{\partial \rm RM}{\partial \theta'_x} \mathrm{cos}\alpha + \frac{\partial \rm RM}{\partial \theta'_y} \mathrm{sin} \alpha \label{eq:direction}$, in which the x'-axis is the direction of proper motion with a constant velocity of $v$.
It is the $bona\ fide$ rotation relationship used in the conversion of plane Cartesian coordinate system.

\subsection{Mapping the projected plasma screen}\label{sec:results}

\begin{figure}
\plotone{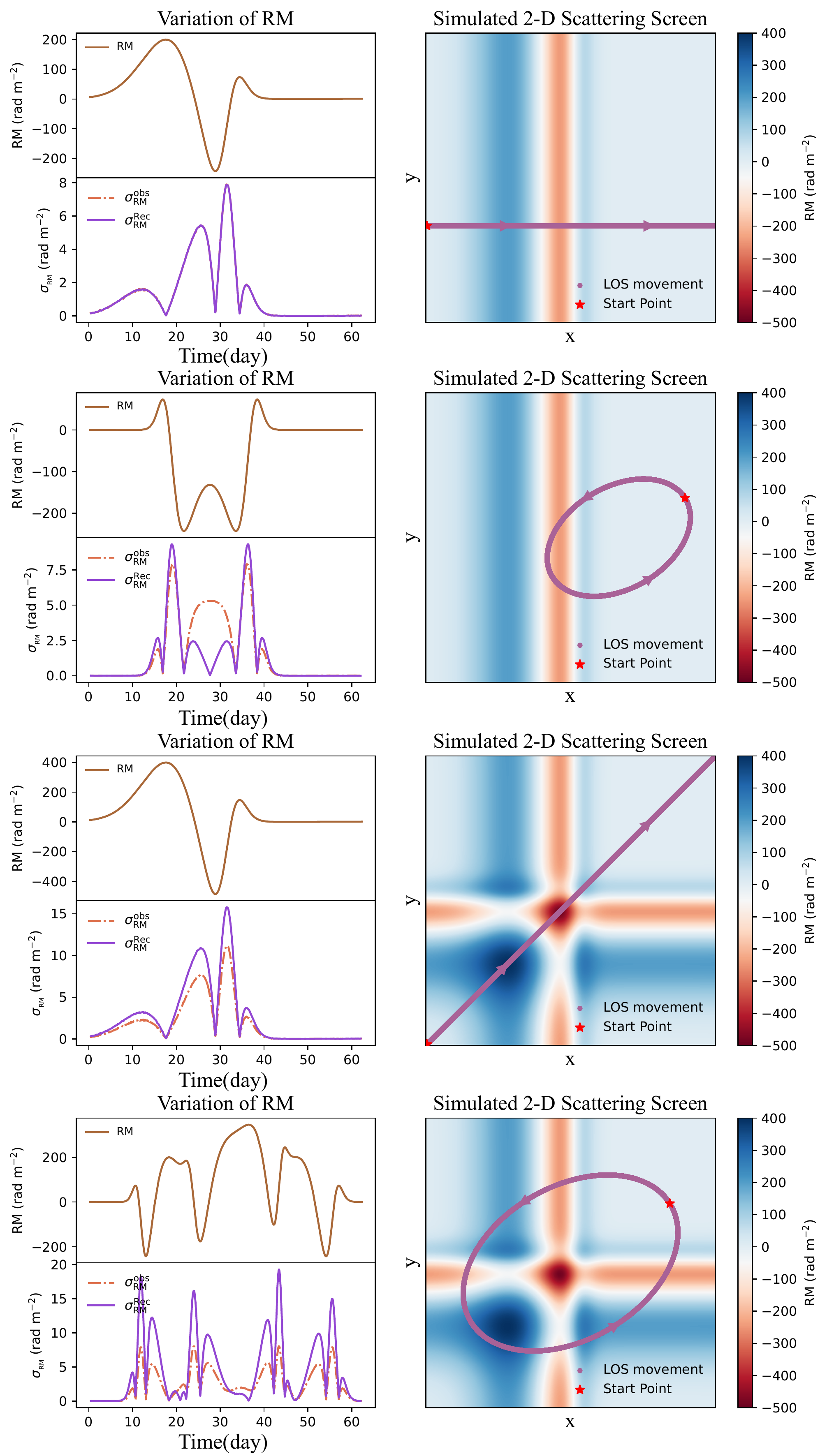}

\caption{Simulation results and their corresponding scattering screen. These four panels share the same contents as follow.
Right: the image of the simulated scattering screen. The purple thick solid line indicates the motion track of the FRB source behind. The red star indicates the start point of the proper motion. The moving direction is anticlockwise.
Left: the sampling result of RM and $\sigma_\mathrm{RM}$ from the simulated screen.  The gold line in the upper panel shows the mean RM sampled from the 2D Gaussian filtering kernel.   The red dashed line  and purple solid line represents the Gaussian weighted 2D $\sigma^\mathrm{obs}_\mathrm{RM}$ and 1D  $\sigma^\mathrm{Rec}_\mathrm{RM}$, respectively.}
\label{fig: screen_total}
\end{figure}

We performed a series of simulations to quantify the effect of the scattering screen with different structures and RM distributions. 
For simplicity, we assumed the FRB source undergoes proper motion with constant velocity. 
A 2D RM screen sized 1500$\times$1500 (unit: hour) was generated using the fitted FRB 20201124A RM template function as is introduced in Section~\ref{sec:simulation}.
The non-evolving RM shown in the latter observation stage was substracted. 
To ensure the sufficient sampling for RM, we generated the motion track of the FRB source at a time step of five minutes. 
Since the majority of observations in \cite{xu22} were performed once per day, we chose the size of the sampling kernel to be 12 hours in order to ensure the Nyquist sampling. 

Figure~\ref{fig: screen_total} shows the simulation results of RM and $\sigma_\mathrm{RM}$ sampled from the foreground scattering screen with different RM configurations and proper motion tracks. 
Divide Figure~\ref{fig: screen_total} into 4 panels in order from top to bottom.
\begin{itemize}
    \item Panel 1 shows the sampled RM and $\sigma_\mathrm{RM}$ with horizontal motion track across the screen that only varies along the x-axis.
The 2D observed standard deviation of RM $\sigma^\mathrm{obs}_\mathrm{RM}$ almost covers the $\sigma^\mathrm{Rec}_\mathrm{RM}$ reconstructed by Gaussian sampling the observed RM curve shown in the upper panel of the left subplot. 
The coincidence can be validated by the sampling method because the 1D sampled $\sigma^\mathrm{Rec}_\mathrm{RM}$ is mathematically the dimensional reduction of 2D sampled $\sigma^\mathrm{obs}_\mathrm{RM}$ once the motion track transverses the x-axis.
    \item The FRB source in panel 2 moves in an elliptical orbit whose symmetric center is coordinated at (1000, 750) with respect to the screen, originated in the lower left. 
The eccentricity is set to $e=0.75$ with an inclination angle $i = 30^\circ$ and semi-major axis $a = 400$ (unit: hrs). 
The orbital elements are selected 
Under the selected orbital elements, the hypothetical FRB source passes through the red region, in accordance with the fitted binary orbit in \cite{wangFY22}, where it crosses the stellar disk near periastron.
\item Panel 3 shows the result sampled along the linear motion track ($y=x$), which resulted in the doubling of sampled RM. Therefore, the $\sigma^\mathrm{Rec}_\mathrm{RM}$ became twice larger than in the panel 1, while the $\sigma^\mathrm{obs}_\mathrm{RM}$ was amplified by a factor of $\sqrt{2}$ . 
The amplification here is consistent with the Equation~\ref{eq:direction} along the moving direction $\alpha = \frac{\pi}{4}$. 
\item Panel 4 shows an elliptical orbit similar with one in panel 2 but longer semi-major axis $a=600$, centered in the geometric center of the canvas
.
The sampled RM and $\sigma_\mathrm{RM}$ variation is strongly modulated by the symmetric RM configuration of the foreground plasma screen. 
\end{itemize}
The correlation between the 2-D observed RM fluctuation time-series $\sigma^\mathrm{obs}_\mathrm{RM}$ and $\sigma^\mathrm{Rec}_\mathrm{RM}$ can be estimated as
\begin{equation}
    r = \frac{\left<\sigma^\mathrm{obs}_\mathrm{RM}\sigma^\mathrm{Rec}_\mathrm{RM}\right>}{\left<(\sigma^\mathrm{obs}_\mathrm{RM})^2\right>} = \left<\sigma^\mathrm{obs}_\mathrm{RM},\sigma^\mathrm{Rec}_\mathrm{RM}\right>.
\end{equation}
By calculating the modified correlation coefficient $r$, the size of reconstructed scattering screen could be inferred at $r=1$. 
With the observing screen size of 6 hrs, the screens reconstructed from the four panels of Figure~\ref{fig: screen_total} are 6.93, 7.97, 4.90, and 4.00 hrs, respectively. 
The reconstructed screen is thus a plausible means to estimate the actual screen size based on the observed RM variation. 

\begin{figure}[htbp]
\plotone{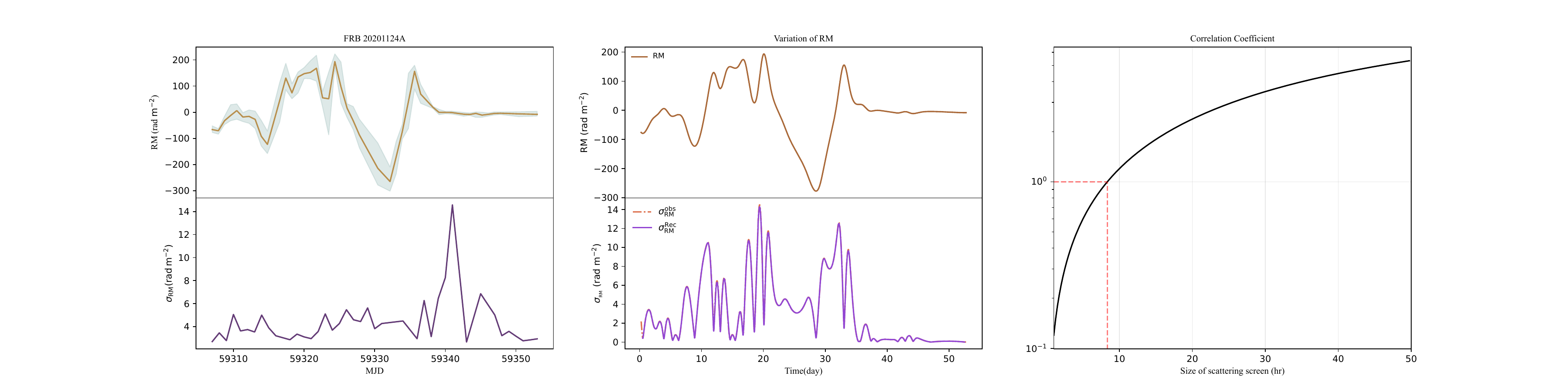}
\caption{The modified correlation coefficient (right) between the $\sigma^\mathrm{Rec}_\mathrm{RM}$ reconstructed from the panel 1 (middle) in Figure~\ref{fig: screen_total} and the observed $\sigma^\mathrm{obs}_\mathrm{RM}$ of FRB 20201124A (left). The characteristic screen size could be obtained when the coefficient $\left<\sigma^\mathrm{obs}_\mathrm{RM}, \sigma^\mathrm{Rec}_\mathrm{RM}\right>$= 1, i.e. the reconstructed RM variation is comparable with the observations.}
\label{fig: coeff}
\end{figure}

We then performed the same simulation using the real observation of RM variation of FRB 20201124A \citep{lu23} as the RM template.
The correlation coefficient between the depolarization variation time-series of FRB 20201124A \citep{lu23} and the reconstructed RM fluctuation $\sigma^\mathrm{Rec}_\mathrm{RM}$ was calculated using screen size trials, plotted in the right subplot of Figure~\ref{fig: coeff}.
and obtained the characteristic screen size  $t_\mathrm{recon} =$8.33 hrs $= \frac{D_\mathrm{B}}{v}\sqrt{\frac{2c\tau_\mathrm{scat}}{D_\mathrm{S}}}$.
Under the same assumption $D_\mathrm{B} = D_\mathrm{S}$, the reconstruction screen distance varying with source velocity appears to be 169 AU at v=100 km/s, lying close to the constraint based on observations. 
Despite the RM variation curve of FRB 20190520B is too coarse to simulate, the reconstructed RM fluctuation comparable with the value listed in Table~\ref{table:info} is also obtained, which resides in $\sim85$ AU.
    
\begin{figure}
\plotone{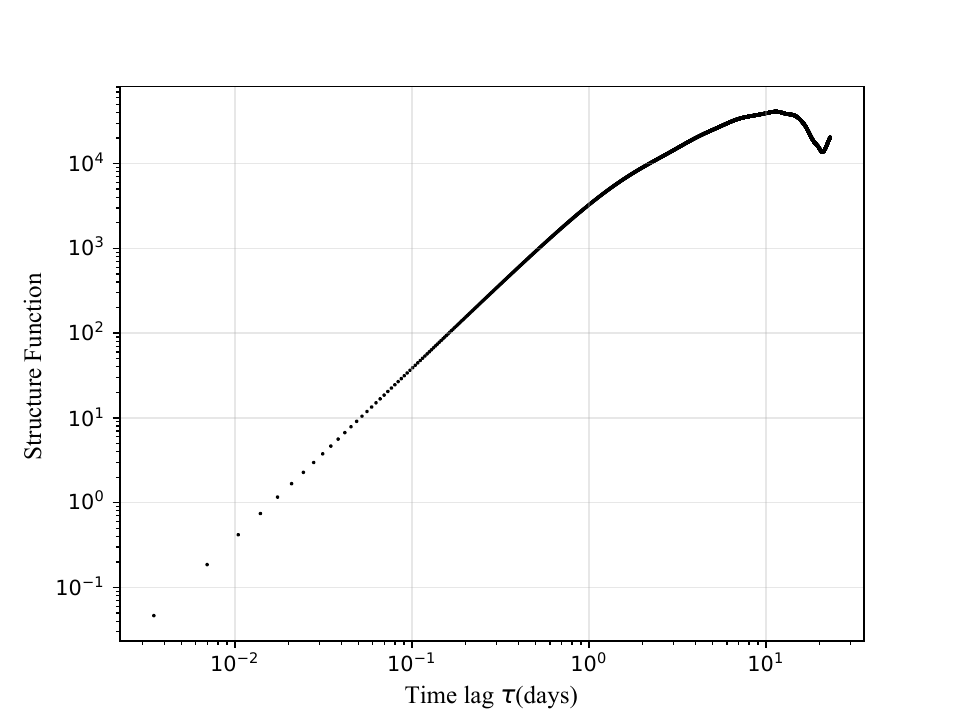}
\caption{Structure function of the interpolated RM variation of FRB 20201124A from March to May 2021. }
\label{fig: SF}
\end{figure}

The first break in the structure function suggests the tentative timescale of RM variation.
Figure~\ref{fig: SF} shows the structure function of the RM of FRB 20201124A, where the first break is adopted as the variation timescale, listed as $\tau_\mathrm{SF}$ in Table~\ref{table:info}.

\section{Observational Overview}\label{sec:obsview}
To provide a simultaneous analysis on propagation effects and its evolution of the selected repeating FRBs, we combined long-term observaions operated by multi-band telescopes.
The observational properties of selected bursts are summarized as follows.

\begin{figure}
\plotone{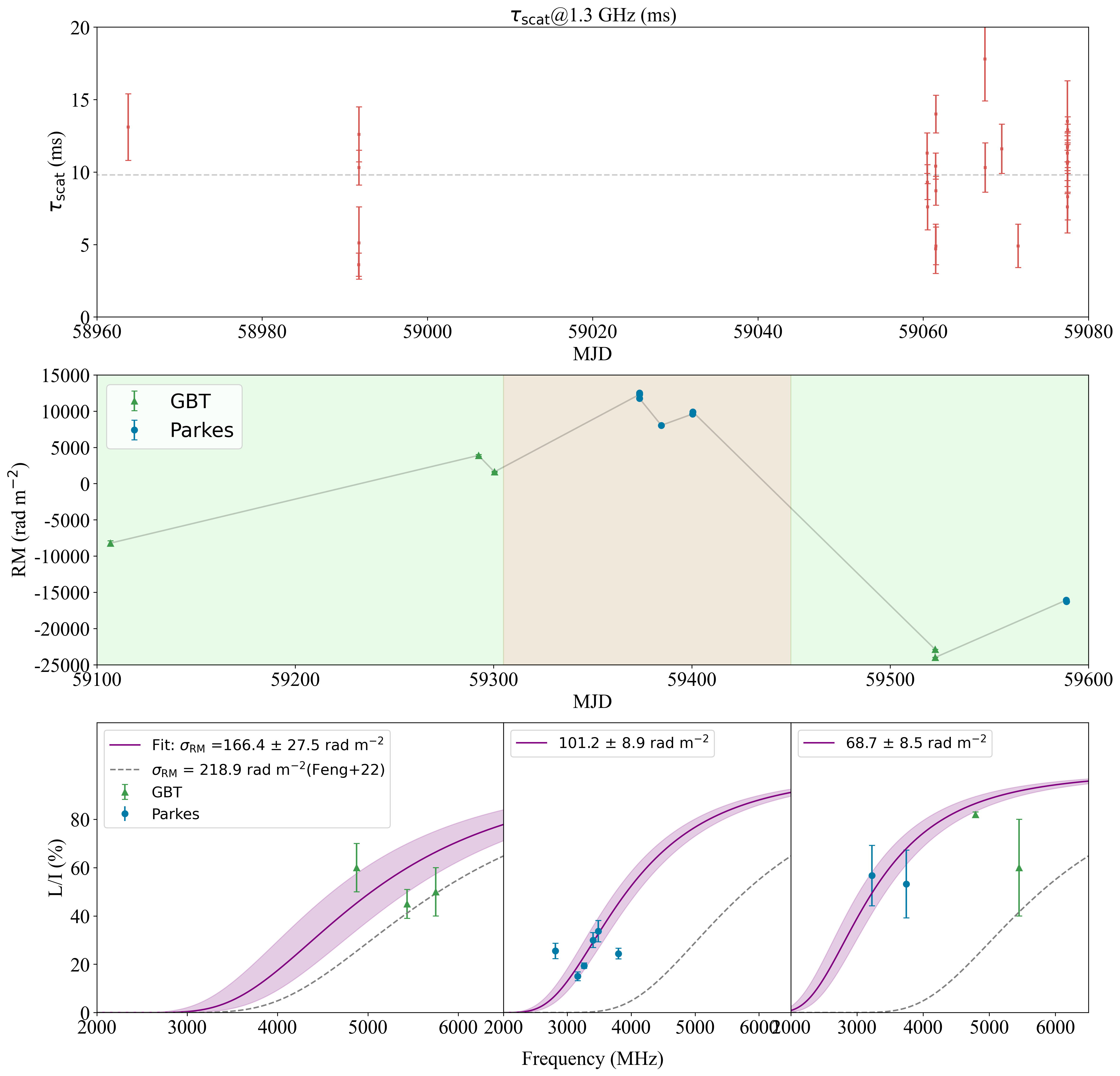}
\caption{The variation of scattering (top) and RM (middle) of FRB 201905202B, and simultaneous depolarization fitting (bottom) in three stages. 
The black dashed lines in the bottom panel represent the fitted depolarization curve by \protect\cite{feng22}, plotted as reference. 
The purple shaded regions indicate the 1$\sigma$ uncertainty of the fitted depolarization curve (purple lines).}
\label{fig: 190520B}
\end{figure}

{\setcitestyle{numbers}
\begin{table*}
    \centering
    \caption{Observational properties including $\tau_\mathrm{scat}$ and $\sigma_\mathrm{RM}$ of some FRB sources}
    \label{table:tau_freq}

    \begin{minipage}{\textwidth}

    \resizebox{0.95\linewidth}{!}{%
    \begin{tabular}{@{}lcccccl@{}}
    \hline
    \hline
    Object 
    & $\nu_\mathrm{depol}$ 
    & $\tau^\mathrm{depol}_\mathrm{scat}$ 
    & $\tau^\mathrm{depol}_\mathrm{sca,YMW16}$ 
    & $\tau^\mathrm{depol}_\mathrm{sca,NE2025}$ 
    & $\tau_\mathrm{SF}(\tau_\mathrm{linear})$ 
    & reference \\
    
    & (GHz) 
    & (ms) 
    & (ms) 
    & (ms) 
    & (day) 
    & \\
    \hline

    FRB 20190520B 
    & $5.0$ 
    & $4.5\times 10^{-2}$ 
    & $1.8\times10^{-6}$ 
    & $6.8\times10^{-7}$ 
    & $183.2$ 
    & \cite{annathomas22,niu22} \\

    FRB 20201124A 
    & $0.88$ 
    & $2.8$ 
    & $0.58$ 
    & $1.1\times 10^{-1}$  
    & $11.33$ 
    & \cite{xu22,201124_sca_2022,lu23} \\

    FRB 20121102A 
    & $2.2$  
    & $<5.5\times 10^{-2}$ 
    & $7.6\times 10^{-2}$ 
    & $1.5\times10^{-2}$ 
    & $3650^{*}$ 
    & \cite{CHIMEcat,121102_decadal} \\

    FRB 20180916B 
    & $0.16$ 
    & $3.8\times 10^1$ 
    & $4.3\times 10^3$ 
    & $2.8\times 10^2$ 
    & $500^{*}$ 
    & \cite{180916_2023,180916_uGMRT_2024,180916_lofar_scatter_2021} \\

    FRB 20190303A 
    & $0.74$ 
    & $1.8$ 
    & $2.3\times10^{-4}$ 
    & $2.0\times10^{-4}$ 
    & $700^{*}$ 
    & \cite{CHIMEcat,CHIMEdepol2023} \\

    FRB 20190417A 
    & $0.89$ 
    & $9.6\times 10^{-1}$ 
    & $1.4\times 10^{-2}$ 
    & $9.2 \times 10^{-4}$  
    & $1200^{*}$ 
    & \cite{CHIMEcat,CHIMEdepol2023} \\
    \hline

    
    \end{tabular}%
    
    }

    \vspace{0.5ex}
    \footnotesize
    \raggedright
    $\nu_\mathrm{depol}$ is the critical frequency at which the linear polarization of an FRB source is depolarized to 50\% at its $\sigma_\mathrm{RM}$. \\
    The scattering timescales, including both observed and model-predicted values, are scaled to their critical depolarization frequency. \\
    $\tau_\mathrm{SF}(\tau_\mathrm{linear})$ is the variation timescale obtained from either the structure function (see Figure~\ref{fig: SF}) or artificial inspection.
    
 
    \end{minipage}
\end{table*}
}
{\it FRB 20190520B}:
\cite{annathomas22} reported the RM variation of FRB 20190520B from 2019 to 2022, which shows dramatic RM reversals over $10^4 $ rad/m$^2$.
We grouped the bursts into three episodes according to the RM variation and fitted the depolarization respectively.
The differential RM changing rate and the depolarization measurements in three stages are shown at the middle and bottom panel in Figure~\ref{fig: 190520B}, respectively.
The scattering timescale was measured by \cite{niu22} 
The depolarization fitting result turns out to be 112.1 $\pm$ 15.0 rad/m$^2$ on average.  
Since the scattering and RM observations were not contemporaneous, we adopted the assumption that scattering timescale maintained stable during the RM variations and plotted in the top panel as reference.

\begin{figure}
\plotone{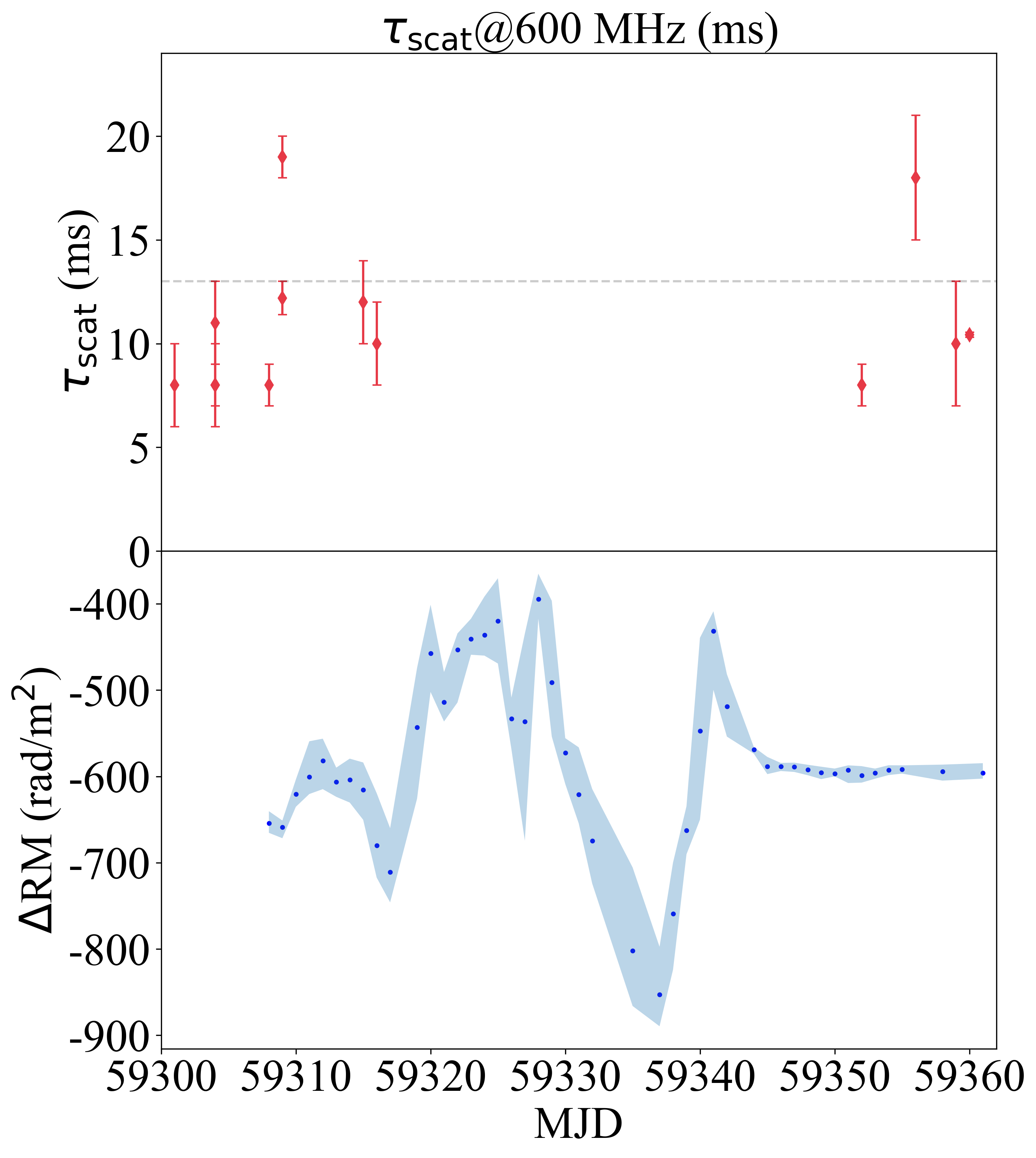}
\caption{The variation of scattering (top) and RM (bottom) of FRB 20201124A from March to May, 2021. }
\label{fig: 201124A}
\end{figure}

{\it FRB 20201124A}: 
The scattering timescale of FRB 20201124A was measured by \cite{201124_sca_2022} centered around 600 MHz, which was roughly stable (8-18 ms) during the observations.
The daily differential RM changing rate and simultaneous depolarization have been measured based on \cite{lu23} utilizing the FAST observations in 2021, where the RM curve is the daily-averaged value.
The simulataneous variations are plotted in Figure~\ref{fig: 201124A}.

\begin{figure}
\plotone{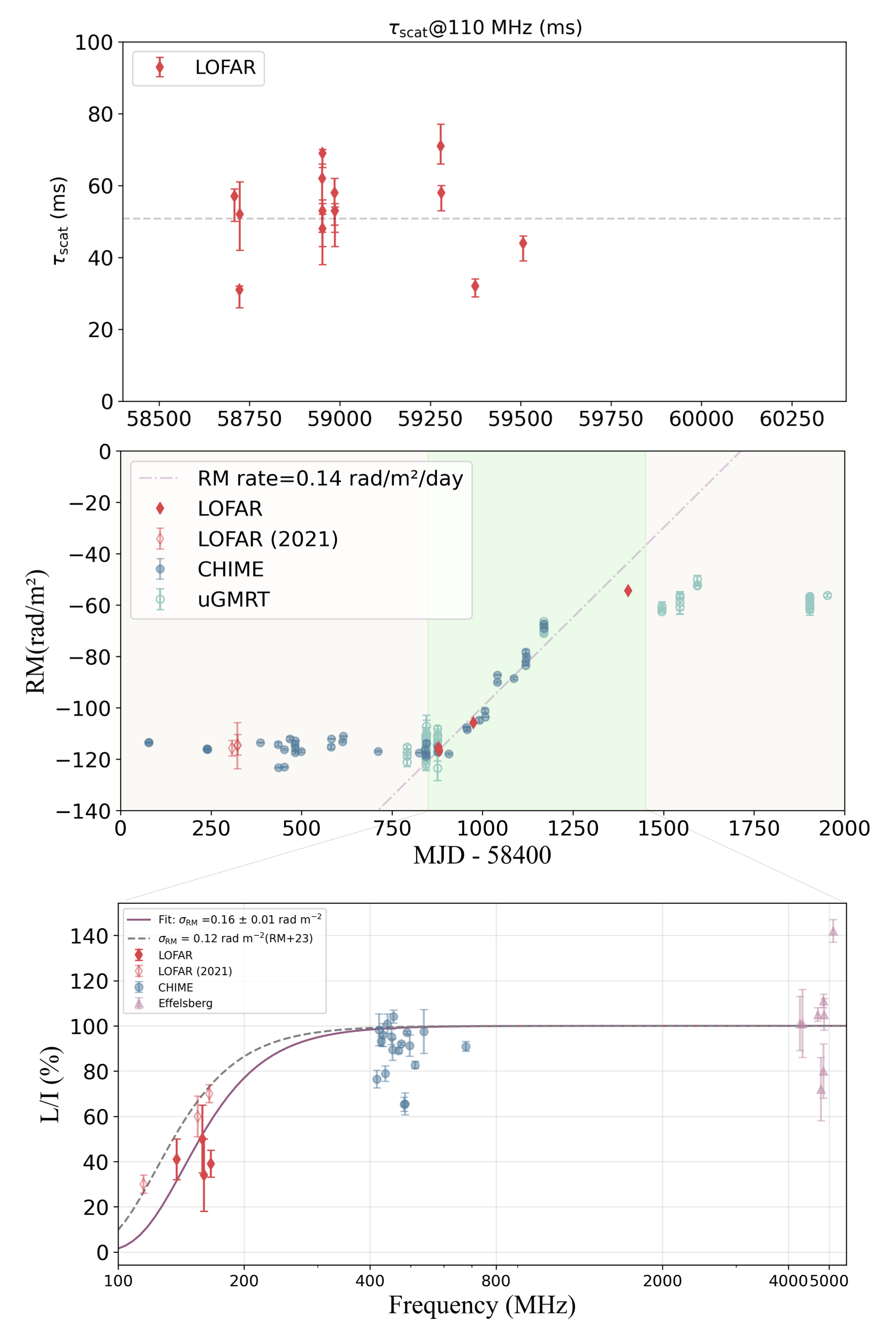}
\caption{The variation of scattering (top) and RM (middle) of FRB 20180916B and simultaneous depolarization fitting (bottom) during the RM decrease phase from April 2021 to June 2022. }
\label{fig: 180916B}
\end{figure}

{\it FRB 20180916B}:  
The absolute value of RM was observed to show a nearly linear decrease trend in 2021 across multiple telescopes covering from 110 MHz to 5 GHz \citep{180916_lofar_scatter_2021,180916_2023,180916_Effel_2023,180916_uGMRT_2024}, suggesting either a dilution of the electron density or a weakening of the line-of-sight magnetic field component.
The depolarization effect was significant below 1 GHz.   
We collected the majority of the bursts that have polarization measurement to figure out the depolarization effect simultaneously with the RM decrease. (full information in Table~\ref{table:bursts}).
For low-frequency observations, the scattering timescale $\tau_\mathrm{scat}$ was measured to be ~50 ms at 150 MHz \citep{180916_lofar_scatter_2021} when the corresponding RM kept constant.
Later simliar scattering timescales were observed by LOFAR during the RM decrease phase \citep{180916_LOFAR_2024_dis}, which implies a relatively stable scattering screen in the local environment of FRB 20180916B.
We fitted the linear decrease rate of RM and the simultaneous depolarization of FRB 20180916B.
The result is shown in Figure~\ref{fig: 180916B}.

{\it FRB 20190303A}:
The RM of FRB 20190303A was observed to show distinct raise-and-fall features revealed by CHIME and FAST sequently. 
The linear fitting of RM variation in two stages turns out to be 0.41 rad/m$^2$/day and -1.18 rad/m$^2$/day, respectively.

\begin{figure}
\plotone{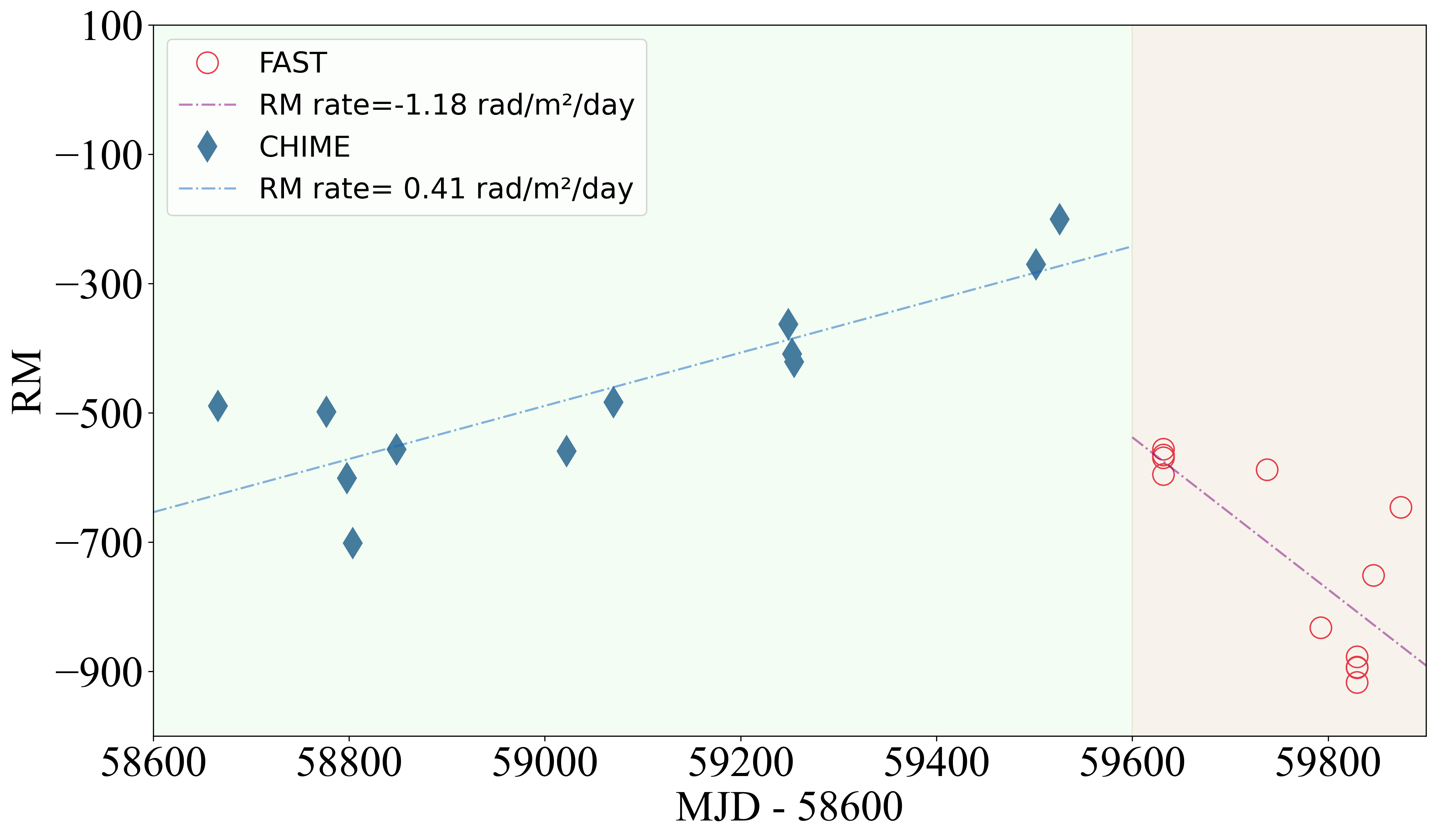}
\caption{The variation of RM of FRB 20190303A. Blue diamonds and red hollow circles represent the FRBs observed by CHIME and FAST, respectively. }
\label{fig: 190303A}
\end{figure}

{\it FRB 20190417A}:  
The RM of FRB 20190417A seemed to fluctuate around 4500 rad/m$^2$ during the FAST observations but remained constant at the same orders of magnitude.
The linear changing rate of RM was fitted throughout all the observations, which turns out to be around 1 rad/m$^2$/day.

\begin{figure}
\plotone{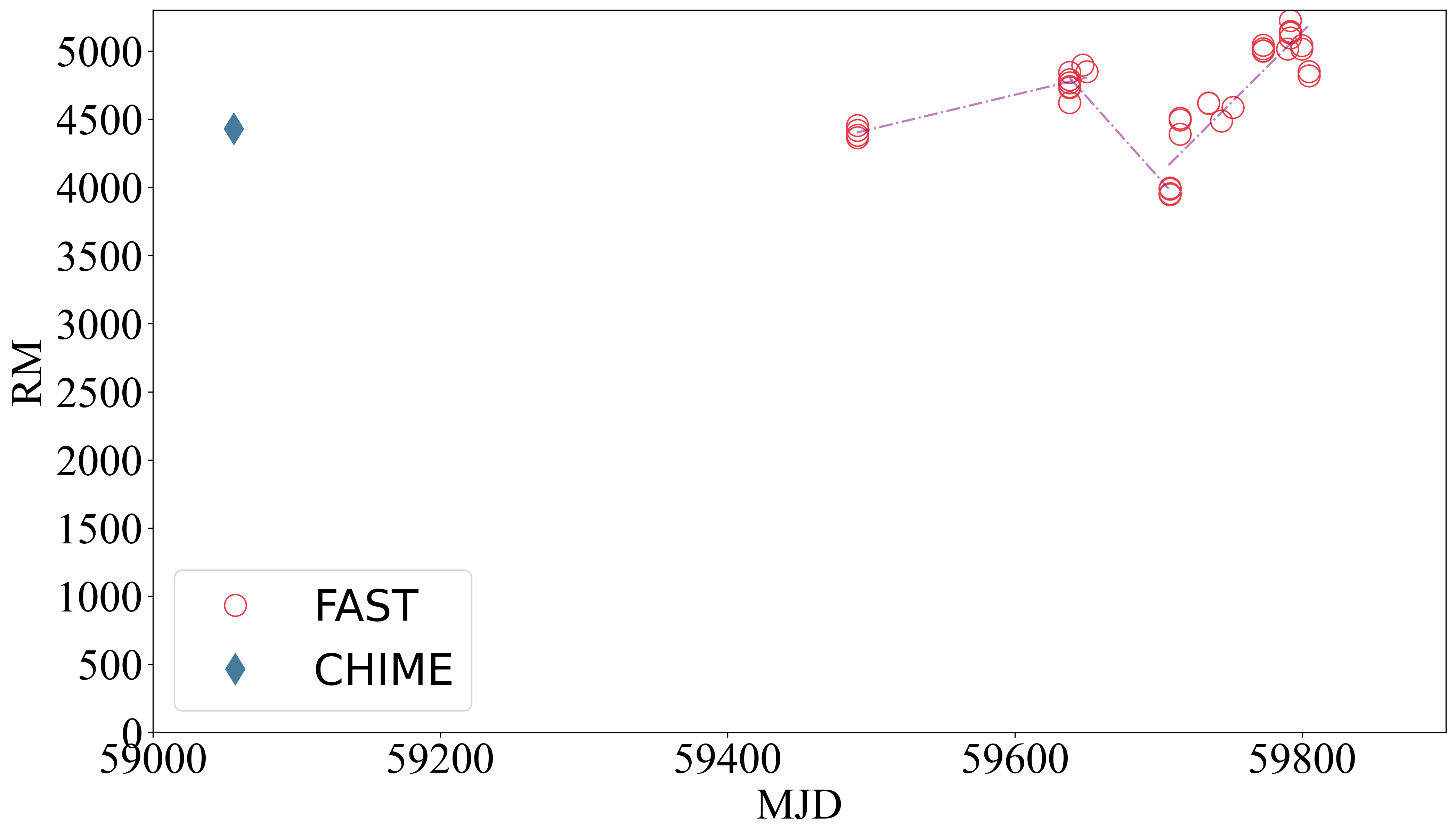}
\caption{The variation of RM of FRB 20190417A. Blue diamond and red hollow circles represent the FRBs observed by CHIME and FAST, respectively. }
\label{fig: 190417A}
\end{figure}

\begin{table}
\centering
\caption{Observational properties of some bursts of FRB 20180916B}
\resizebox{\columnwidth}{!}{%
\begin{tabular}{lrrrccccc}
\toprule
        Number & MJD & Frequency & RM  & L/I \\
        - & - & (MHz) & $\rm rad/m^2$ & - \\
\midrule
\hline
        \hline
         &  & LOFAR &  &  \\
        \hline
        1 & 58708.14344 & 165 & -115.71(-) & 0.70(4) \\
        2 & 58722.20383 & 155 & -114.78(-) & 0.60(9) \\
        3 & 58723.21829 & 115 & -114.43(-) & 0.30(4) \\
        4 & 59278.71223 & 166.6 & -115.5(-) & 0.39(6) \\
        5 & 59280.45791 & 137.95 & -116.6(-) & 0.41(9) \\
        6 & 59374.26540 & 160.3 & -105.8(-) & 0.34(16) \\
        7 & 59802.43979 & 158.95 & -54.4(-) & 0.50(15) \\
        \hline
        &  & CHIME &  &  \\ 
        \hline
        1 & 58477.16185 & 450.0 & -113.59(28) & 0.897(20) \\
        2 & 58478.15521 & 550.0 & -113.43(14) & 0.873(16) \\
        3 & 58638.71643 & 625.0 & -115.99(12) & 0.9773(84) \\
        4 & 58639.70561 & 490.0 & -115.98(17) & 0.991(20) \\
        5 & 58639.71008 & 480.0 & -116.22(41) & 0.944(15) \\
        6 & 58786.32075 & 440.0 & -113.6(1.3) & 0.880(46) \\
        7 & 58835.17324 & 480.0 & -114.24(79) & 0.743(32) \\
        8 & 58836.17198 & 425.0 & -123.3(2.3) & 1.002(83) \\
        9 & 58852.13628 & 695.0 & -123.0(2.6) & 0.761(64) \\
        10 & 58852.13773 & 487.5 & -116.18(36) & 0.848(27) \\
        11 & 58868.07586 & 607.5 & -112.05(83) & 0.745(26) \\
        12 & 58882.04681 & 707.5 & -114.4(2.3) & 0.933(78) \\
        13 & 58883.03995 & 435.0 & -117.40(64) & 0.941(41) \\
        14 & 58883.04405 & 687.5 & -112.8(2.5) & 1.031(64) \\
        15 & 58883.05372 & 672.5 & -115.85(91) & 0.636(32) \\
        16 & 58899.00706 & 475.0 & -116.89(51) & 0.892(33) \\
        17 & 58981.77662 & 725.0 & -115.22(84) & 0.933(26) \\
        18 & 58982.77157 & 440.0 & -112.1(1.2) & 0.877(60) \\
        19 & 59013.69287 & 450.0 & -113.19(31) & 0.932(27) \\
        20 & 59014.68533 & 610.0 & -110.97(14) & 0.938(22) \\
        21 & 59111.40643 & 485.0 & -116.9(3.1) & 0.443(39) \\
        22 & 59225.10428 & 525.0 & -117.5(1.3) & 0.908(78) \\
        23 & 59241.06071 & 640.0 & -116.40(92) & 0.911(71) \\
        24 & 59244.04400 & 485.0 & -118.87(63) & 0.884(37) \\
        25 & 59244.06229 & 550.0 & -117.97(64) & 0.782(25) \\
        26 & 59245.05833 & 455.0 & -113.75(87) & 1.011(34) \\
        27 & 59275.96877 & 455.0 & -115.14(23) & 1.041(28) \\
        28 & 59276.96257 & 452.5 & -117.20(68) & 0.895(47) \\
        29 & 59277.96034 & 450.0 & -116.66(28) & 0.951(34) \\
        30 & 59306.89010 & 677.5 & -117.94(34) & 0.909(21) \\
        31 & 59355.74759 & 497.5 & -107.6(1.7) & 0.912(47) \\
        32 & 59357.74680 & 440.0 & -108.42(78) & 1.009(43) \\
        33 & 59390.65203 & 430.0 & -104.86(97) & 0.959(51) \\
        34 & 59406.60025 & 482.5 & -101.15(77) & 0.653(31) \\
        35 & 59407.59936 & 475.0 & -103.624(25) & 0.9205(63) \\
        36 & 59440.50726 & 485.0 & -87.2(1.5) & 0.655(48) \\
        37 & 59440.51585 & 435.0 & -90.07(44) & 0.789(34) \\
        38 & 59486.38963 & 537.5 & -88.63(62) & 0.975(97) \\
        39 & 59519.30410 & 420.0 & -78.25(91) & 0.982(70) \\
        40 & 59519.30470 & 490.0 & -83.544(42) & 0.9702(66) \\
        41 & 59519.30909 & 512.5 & -81.93(19) & 0.827(16) \\
        42 & 59521.28871 & 425.0 & -80.08(26) & 0.932(22) \\
        43 & 59569.16266 & 415.0 & -67.3(1.1) & 0.764(39) \\
        44 & 59570.15498 & 467.5 & -69.032(88) & 0.891(12) \\
        \hline 
\bottomrule
\end{tabular}
}
\label{table:bursts}
\end{table}

\begin{table}
\centering
\caption{Continuous}
\resizebox{\columnwidth}{!}{%
\begin{tabular}{lrrrccccc}
\toprule
        Number & MJD & Frequency & RM  & L/I \\
        - & - & (MHz) & $\rm rad/m^2$ & - \\
\midrule
\hline
        \hline
        & & {Effelsburg} & & \\
        \hline
        1 & 59353.80212 & 4782.6 & - & 0.72(14) \\
        2 & 59353.83006 & 4704.6 & - & 1.05(3) \\
        3 & 59435.75198 & 4851.0 & - & 0.80(12) \\
        4 & 59484.35741 & 5117.6 & - & 1.42(5) \\
        5 & 59484.38322 & 4867.6 & - & 1.05(7) \\
        6 & 59484.44951 & 4259.9 & - & 1.01(12) \\
        7 & 59484.45820 & 4852.0 & - & 1.11(3) \\
        8 & 59484.46443 & 4317.55& -  & 1.01(15) \\
        \hline
        & & {uGMRT} & & \\
        \hline
        1 & 59191.82861 & 678.68 & -121.12(1.74) & - \\
        2 & 59191.83428 & 656.32 & -118.8(1.45) & - \\
        3 & 59191.84099 & 660.64 & -117.2(0.90) & - \\
        4 & 59191.85129 & 635.93 & -115.22(0.68) & - \\
        5 & 59243.40879 & 601.03 & -111.34(3.50) & - \\
        6 & 59243.41312 & 595.54 & -116.39(1.38) & - \\
        8 & 59243.42712 & 623.77 & -116.48(3.70) & - \\
        9 & 59243.43115 & 683.38 & -110.58(1.85) & - \\
        10 & 59243.43648 & 666.13 & -115.78(0.69) & - \\
        12 & 59243.45379 & 711.62 & -110.8(0.83) & - \\
        13 & 59243.45408 & 659.46 & -115.03(0.85) & - \\
        14 & 59243.45438 & 603.77 & -120.03(3.08) & - \\
        15 & 59243.45525 & 672.40 & -111.83(0.36) & - \\
        17 & 59243.45785 & 643.77 & -115.27(0.51) & - \\
        18 & 59243.46044 & 586.13 & -117.59(0.86) & - \\
        19 & 59243.46044 & 686.91 & -112.51(1.02) & - \\
        20 & 59243.46139 & 662.60 & -121.82(2.08) & - \\
        23 & 59243.48092 & 629.26 & -117.04(2.83) & - \\
        24 & 59243.48160 & 602.21 & -116.12(2.35) & - \\
        25 & 59243.48160 & 598.68 & -119.32(2.15) & - \\
        26 & 59243.48198 & 591.62 & -117.2(1.46) & - \\
        27 & 59243.48233 & 657.50 & -114.95(0.14) & - \\
        29 & 59243.51028 & 657.11 & -113.63(1.28) & - \\
        30 & 59243.51028 & 631.62 & -113.25(1.92) & - \\
        31 & 59243.52054 & 612.79 & -117.07(2.18) & - \\
        32 & 59243.52054 & 624.95 & -117.13(2.79) & - \\
        33 & 59243.54053 & 654.36 & -114.48(0.18) & - \\
        34 & 59243.54816 & 657.89 & -113.18(0.17) & - \\
        36 & 59243.55634 & 659.46 & -115.51(0.20) & - \\
        37 & 59243.55646 & 652.40 & -114.83(3.17) & - \\
        38 & 59243.56064 & 698.68 & -111.18(0.67) & - \\
        39 & 59243.56652 & 692.40 & -110.5(0.40) & - \\
        40 & 59243.56786 & 648.48 & -115.49(0.17) & - \\
        41 & 59243.56880 & 606.52 & -115.84(4.03) & - \\
        42 & 59243.57030 & 684.17 & -111.87(1.26) & - \\
        43 & 59243.57629 & 670.05 & -114.49(1.26) & - \\
        44 & 59243.57750 & 635.54 & -117.06(1.70) & - \\
        45 & 59243.57762 & 573.58 & -120.61(3.84) & - \\
        46 & 59244.59722 & 641.81 & -113.54(0.25) & - \\
        47 & 59244.59940 & 570.44 & -110.08(2.47) & - \\
        48 & 59244.60563 & 616.72 & -111.02(8.15) & - \\
        49 & 59244.61136 & 582.99 & -107.18(2.33) & - \\
        50 & 59274.58289 & 629.26 & -114.31(0.83) & - \\
        \hline 
\bottomrule
\end{tabular}
}
\end{table}

\begin{table}
\centering
\caption{Continuous}
\resizebox{\columnwidth}{!}{%
\begin{tabular}{lrrrccccc}
\toprule
        Number & MJD & Frequency & RM  & L/I \\
        - & - & (MHz) & $\rm rad/m^2$ & - \\
\midrule
\hline
        \hline
         51 & 59275.47513 & 587.30 & -108.12(0.99) & - \\
        52 & 59275.48829 & 661.42 & -112.77(0.24) & - \\
        53 & 59275.49104 & 614.36 & -116.97(3.71) & - \\
        54 & 59275.51090 & 680.76 & -123.44(4.86) & - \\
        55 & 59275.51500 & 603.38 & -110.63(0.31) & - \\
        56 & 59275.52895 & 612.79 & -110.71(2.38) & - \\
        57 & 59275.53099 & 616.72 & -111.39(1.74) & - \\
        58 & 59568.73872 & 591.62 & -71.04(1.10) & - \\
        59 & 59568.74316 & 664.95 & -66.29(0.49) & - \\
        60 & 59568.75117 & 630.05 & -68.45(0.26) & - \\
        62 & 59568.75264 & 649.26 & -67.61(0.16) & - \\
        64 & 59568.75326 & 586.52 & -69.56(0.67) & - \\
        65 & 59568.76806 & 635.15 & -70.44(0.95) & - \\
        66 & 59568.76958 & 634.75 & -69.48(0.58) & - \\
        68 & 59894.79429 & 692.40 & -60.53(1.77) & - \\
        69 & 59894.79637 & 649.66 & -62.47(0.24) & - \\
        70 & 59894.84244 & 659.07 & -60.41(0.60) & - \\
        71 & 59894.84801 & 652.79 & -61.74(0.35) & - \\
        72 & 59944.56908 & 663.77 & -60.68(2.92) & - \\
        73 & 59944.60543 & 587.30 & -58.98(3.12) & - \\
        74 & 59944.63033 & 645.74 & -59.06(4.33) & - \\
        75 & 59944.64854 & 644.95 & -57.14(0.09) & - \\
        76 & 59944.67617 & 655.15 & -56.34(0.96) & - \\
        79 & 59993.48844 & 661.03 & -52.44(0.28) & - \\
        83 & 59993.53836 & 628.48 & -52.13(0.25) & - \\
        85 & 59993.56582 & 601.42 & -50.04(1.56) & - \\
        88 & 60303.40070 & 590.05 & -59.12(2.07) & - \\
        89 & 60303.40274 & 602.60 & -60.28(2.19) & - \\
        91 & 60303.44708 & 653.58 & -56.54(0.17) & - \\
        92 & 60303.44867 & 652.79 & -57.7(0.13) & - \\
        93 & 60303.45043 & 669.26 & -57.12(0.72) & - \\
        95 & 60303.48179 & 648.48 & -61.02(2.10) & - \\
        96 & 60303.48308 & 665.34 & -57.49(0.51) & - \\
        98 & 60303.48435 & 599.85 & -60.39(2.63) & - \\
        99 & 60303.48435 & 601.03 & -61.65(2.33) & - \\
        100 & 60352.34462 & 652.40 & -56.22(0.69) & - \\
        \hline 
\bottomrule
\end{tabular}
}
\end{table}

\section{Discussion on Parameter Uncertainty and Robustness}\label{sec:uncertainty}

The estimated $D_\mathrm{S}$ relies on the assumption that the RM variation, depolarization, and scattering originate from the same foreground screen.
\cite{180916_LOFAR_2024_dis} proposed that the uncorrelated depolarization and scattering may arise from different regions, which is consequently not suitable for the precondition of Equation~\ref{eq:dist}. 
Besides, the scattering timescale in these formulae is the component contributed by the local scattering screen, which may be far less than the observed scattering applied in this work. 
Assuming that the scattering of these FRBs is dominated by the Galactic interstellar medium, we derived the Galactic scattering timescales for each FRB source (listed in Table~\ref{table:info}) by adopting their Galactic DM contribution $\mathrm{DM}_\mathrm{Gal}$ using both the YMW16 \citep{ymw16} and NE2025 \citep{NE2025} Galactic electron density models.
NE2025 considered several Galactic clumps and voids to estimate the electron density and its fluctuation for specific sightlines. 
The YMW16-based Galactic scattering estimates primarily rely on the empirical $\tau_\mathrm{scat}-\mathrm{DM}$  relation derived from 128 Galactic pulsars observed at 327 MHz \citep{Krishnakumar2015}. 
The observed scattering tails in several FRBs deviate from model predictions, suggesting a non-Galactic origin—-likely arising from their host galaxies or local environments rather than the Milky Way. 
However, the apparent agreement between YMW16 predictions and the scattering of  FRB 20121102A and FRB 20180916B implies that the Galactic dominance on the scattering could not be ruled out.
The assumption that the scattering maintains stable across the observations, may also be incomplete in some circumstances. 

Apart from the apparent inconsistencies between the theoretical assumption and the observations, we note that the observational parameters used to estimate the $D_\mathrm{S}$ may show bias.
The measurements on the RM changing rate are based on daily average RM variation curve for acitve FRB repeaters, and all the bursts with RM for other FRB repeaters (such as FRB 20190303A and FRB 20190417A) to maximize the data-sets.
The variation timescale is either estimated from the structure function or the time span selected from artificial inspection, which may not represent the intrinsic variation timescale of the surrounding environment.
The discrepancy between the selected variation timescale ant the exact time span of RM variation is negligible to the uncertainties of the proper motion velocity.

Besides, $D_\mathrm{S}$ may vary with time in some specific circumstances.
For instance, $D_\mathrm{S}$ in the binary scenario could be correlated with the variation timescale according to the Kepler's Third Law, if the RM variation is caused by the periodic orbital motion. 
However, RM variations generally happens as long as the FRB source moves transversely behind inhomogeneous magneto-active screen as described in Section~\ref{sec:Geometry}, despite of the dynamic evolution of the foreground screen.
Therefore, the time dependence of $D_\mathrm{S}$ was not included in case of these unclear physical condition.

\section{Environmental Models}\label{sec:models}

\subsection{Binary}\label{sec:binary}
The RM variations of some repeating FRBs can be explained by a high-eccentricity massive binary system consisting of a neutron star and a massive star \citep{wangFY22,Zhao2023}. For elliptical orbits, the source-to-screen distance must be corrected from $a(1-e)$ to $a(1+e)$, where $e$ is the eccentricity. 

The evolution of DM and RM contributed by the binary system is mainly determined by the characteristic value DM$_0$ and RM$_0$ determined by the stellar wind and the modulation of orbital geometry. The influence of orbital geometry can be found in \citealt{wangFY22,Zhao2023}. For simplicity, it is not considered in detail in this work. The electron density of the stellar wind is given by
\begin{equation}
n_{\mathrm{w}}(r)=n_{\mathrm{w}, 0}\left(\frac{r}{R_{\star}}\right)^{-2},
\end{equation}
where $R_{\star}$ is the radius of the star. The density at the surface of the massive star is $n_{\mathrm{w}, 0}=\dot{M} / 4 \pi R_{\star}^2 v_{\mathrm{w}} \mu_{\mathrm{i}} m_{\mathrm{p}}$, where $\dot{M}$ is the mass-loss rate, $v_{\mathrm{w}}$ is the wind velocity, $\mu_{\mathrm{i}}\simeq 1.29$ is the mean ion molecular weight \citep{Dubus2013} and $m_{\mathrm{p}}$ is the mass of the protons. The DM$_0$ contributed by the stellar wind is
\begin{equation}
\begin{aligned}
\mathrm{DM_{w, 0}}&\simeq n_{\mathrm{w}}(a) \cdot a= \frac{\dot{M}}{4 \pi v_\mathrm{w} \mu_\mathrm{i} m_\mathrm{p} a^2}\cdot a \\
&\approx  5  ~\mathrm{p c~cm}^{-3}v_\mathrm{w,8}^{-1}\left(\frac{\dot{M}}{10^{-8} ~M_\odot \mathrm{yr}^{-1}}\right)\left(\frac{a}{1~\mathrm{AU}}\right)^{-1}.               
\end{aligned}
\end{equation}

The large-scale magnetic field in the wind is $B(r)\sim B_0(r/R_{\star})^{-\beta}$, where $B_0$ is the magnetic field strength. The slope is $\beta=2$ for a radial field and $\beta=1$ for a toroidal field. The RM$_0$ contributed by the stellar wind for a toroidal field is 
\begin{equation}
\begin{aligned}
\mathrm{RM}_0 \sim & \frac{e^3 B n_\mathrm{w} a}{2 \pi m_e^2 c^4} \simeq 2 \times 10^4 \mathrm{~rad} \mathrm{~m}^{-2}\left(\frac{B_0}{1 ~\mathrm{G}}\right)\left(\frac{R_\star}{1~ R_{\odot}}\right) \\    
& \times\left(\frac{\dot{M}}{10^{-8} M_{\odot} \mathrm{~yr}^{-1}}\right)\left(\frac{a}{1 \mathrm{~AU}}\right)^{-2}\left(\frac{v_w}{10^3 \mathrm{~km} \mathrm{~s}^{-1}}\right)^{-1},
\end{aligned}
\end{equation}
and for a radial field is
\begin{equation}
\begin{aligned}
\mathrm{RM}_0 \sim & \frac{e^3 B n_\mathrm{w} a}{2 \pi m_e^2 c^4} \simeq 94 \mathrm{~rad} \mathrm{~m}^{-2}\left(\frac{B_0}{1 ~\mathrm{G}}\right)\left(\frac{R_\star}{1~ R_{\odot}}\right) \\    
& \times\left(\frac{\dot{M}}{10^{-8} M_{\odot} \mathrm{~yr}^{-1}}\right)\left(\frac{a}{1 \mathrm{~AU}}\right)^{-3}\left(\frac{v_w}{10^3 \mathrm{~km} \mathrm{~s}^{-1}}\right)^{-1}.
\end{aligned}
\end{equation}

\cite{Zhao2023} studied the RM contributed by clumps in stellar wind. However, the traversing timescale of clumps in a binary system is $<1$ d, which may lead to a short-time scale stochastic variation of RM ($\delta \mathrm{RM}\sim0.06-8$ rad m$^{-2}$), but fails to account for the long-time scale variation. The RM scatter contributed by the clumps in the stellar wind is \citep{yang22}
\begin{equation}
\begin{aligned}
\sigma_{\mathrm{RM}} & \simeq\left(\frac{\Delta R}{l_{\mathrm{s}}}\right)^{1 / 2} \delta \mathrm{RM}\left(l_{\mathrm{s}}\right) \\
& \simeq 2.6 ~\mathrm{rad~m^{-2}}\left(\frac{\Delta R}{10~R_\odot}\right)^{1 / 2}\left(\frac{l_{\mathrm{s}}}{10^{11}~\mathrm{cm}}\right)^{-1 / 2}\left(\frac{\delta \mathrm{RM}}{1~\mathrm{rad~m^{-2}}}\right),
\end{aligned}
\end{equation}
where $\Delta R$ is the thickness of the magnetized plasma screen and $l_{\mathrm{s}}\sim 10^{10}-10^{11}$ cm is the size of clumps in the stellar wind \citep{Zhao2023}.

\subsection{Supernova remnant}\label{sec:snr}

Supernova explosion kick velocity of a newborn NS is up to $V_\mathrm{k}\sim 10^3$ km/s \citep{transV2012}. Only when $V_\mathrm{k}>V_{\mathrm{sh}}$, the evolution of RM is dominated by the FRB source's proper motion rather than the evolution of SNRs. This is possible for the older SNR, whose age can be estimated as
\begin{equation}\label{eq:t_age}
t_{\mathrm{age}}>7.1\times10^3 \mathrm{~yr} \left(\frac{n_0}{0.1 \mathrm{~cm}^{-3}}\right)^{-1 / 3}\left(\frac{E}{10^{51} \mathrm{erg}}\right)^{1 / 3}\left(\frac{V_\mathrm{k}}{1000 \mathrm{~km~s^{-1}}}\right)^{-5/3}.
\end{equation}
If the kick velocity is smaller (e.g. 10-1000 km/s, \citealt{Hansen1997}), then the age given by equation (\ref{eq:t_age}) may already exceed the active age of the magnetar $t_{\mathrm{act}}\sim10^4$ yr. The source-to-screen distance can be roughly estimated as the size of the SNR $D_\mathrm{s}\sim R_{\mathrm{sh}}$.

The magnetic field in the shocked region is
\begin{equation}\label{eq:Bsh}
\frac{B^2}{8 \pi}=\epsilon_{\mathrm{B}} u_{\mathrm{th}},
\end{equation}
where $\epsilon_{\mathrm{B}}$ is the magnetic energy density fraction and $u_{\mathrm{th}}=9 \rho_{0} V_\mathrm{sh}^2 / 8$. The RM from the shocked shell is
\begin{equation}
\begin{aligned}
 \mathrm{RM_{sh}} &\simeq8.1 \times 10^5 \mathrm{rad~m^{-2}} \left(\frac{4n_0}{\mathrm{~cm}^{-3}}\right) \left(\frac{B_\mathrm{sh}}{\mathrm{G}}\right) \left(\frac{\xi R_{\mathrm{sh}}}{\mathrm{pc}}\right)\\
& \simeq 16 \mathrm{~rad} \mathrm{~m}^{-2}\epsilon_{\mathrm{B},-3}^{1/2} \xi_{-1}\left(\frac{n_0}{0.1 \mathrm{~cm}^{-3}}\right)^{1.1}\left(\frac{E}{10^{51} \mathrm{erg}}\right)^{2 / 5}\left(\frac{t}{7000 \mathrm{~yr}}\right)^{-1/ 5},
\end{aligned}
\end{equation}
where $\xi$ is the thickness of the shock. 
The RM associated with the shocked shell exhibits a gradual evolution for older SNRs ($\mathrm{RM}\propto t^{-1/5}$). 
The observed RM variation timescale is a few years, and the RM fluctuation is 
\begin{equation}
\left | \frac{\delta \mathrm{RM}}{\mathrm{RM}} \right |\sim \frac{t_{\mathrm{obs}}}{5t_{\mathrm{age}}}=2\times10^{-4}\left(\frac{t_{\mathrm{obs}}}{1\mathrm{~yr}}\right)\left(\frac{t_{\mathrm{age}}}{1000 \mathrm{~yr}}\right)^{-1}.       
\end{equation}\label{eq:RMfrac_SNR}

\begin{figure}
\plotone{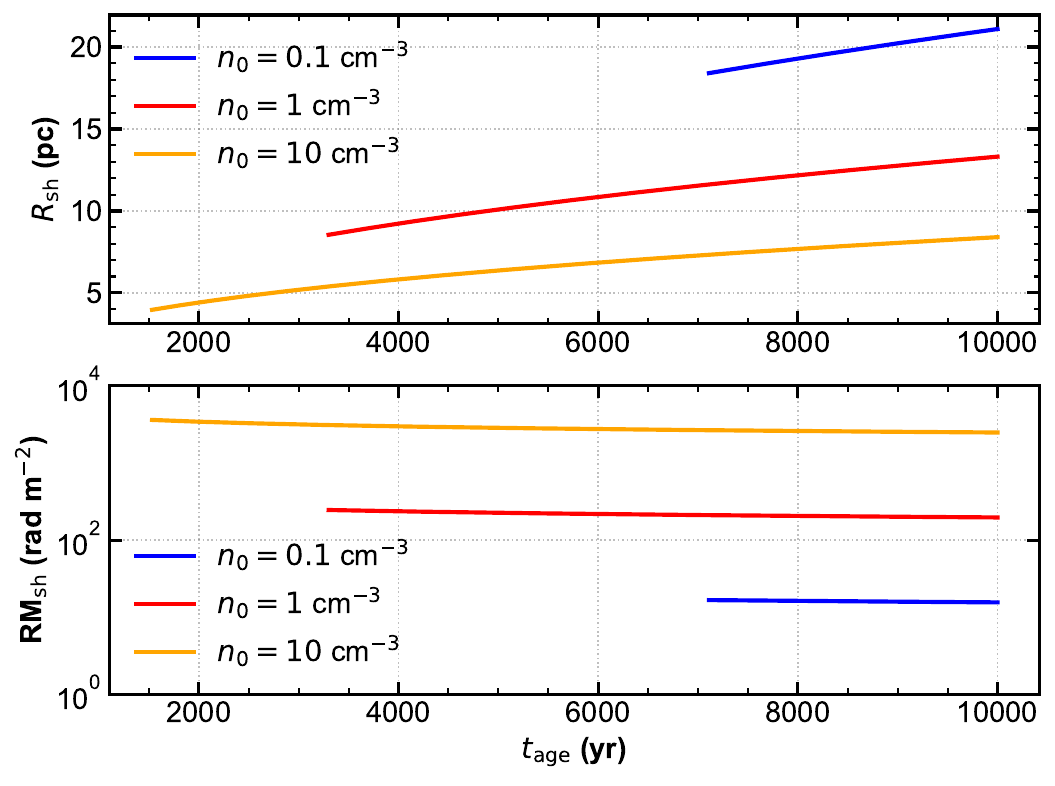}
\caption{The evolution of the radius of shocked sheell of SNR and its RM contribution, varying as a function of electron density.}
\label{fig:RM_SNR}
\end{figure}

For the ambient medium with the density $n_0=0.1-10$ cm$^{-3}$, the SNR radius and RM from the shocked shell are shown in Figure \ref{fig:RM_SNR}. 
The minimum SNR age is determined by Equation (\ref{eq:t_age}), and the maximum age is taken as the active timescale of the magnetar ($t_{\mathrm{act}}\sim10^4$ yr), so the size of the SNR is  $\sim 5-20$ pc.

\subsection{Clumps in SNR}\label{sec:clumps}

The RM variations of FRBs could also be attributed to filaments in SNRs across the line of sight \citep{Katz2021}. From the X-ray observations of SNRs, the size of clumps is $D_{\mathrm{c}}\sim 10^{14}-10^{16}$ cm and the magnetic field strength is $B_{\mathrm{c}}\sim 0.1-0.3 \mathrm{mG}$ \citep{Patnaude2009,Fesen2011}. The traversing timescale of the clump is 
\begin{equation}
    t_{\mathrm{s}}=\frac{D_{\mathrm{c}}}{V_\mathrm{k}}\simeq 3.2 \mathrm{~yr}\left(\frac{D_{\mathrm{c}}}{10^{16}\mathrm{~cm}}\right)\left(\frac{V_\mathrm{k}}{1000 \mathrm{~km~s^{-1}}}\right)^{-1},
\end{equation}\label{eq:t_clumps}
which is consistent with the RM variation timescale of repeating FRBs. The RM contributed by clumps in the SNR can be estimated by
\begin{equation}
    \mathrm{RM_{c}} \simeq 26\mathrm{~rad} \mathrm{~m}^{-2} \chi_3 \left(\frac{n_0}{0.1 \mathrm{~cm}^{-3}}\right)\left(\frac{B_\mathrm{c}}{0.1\mathrm{~mG}}\right)\left(\frac{D_{\mathrm{c}}}{10^{16}\mathrm{~cm}}\right),  
\end{equation}
where $\chi\sim10^2-10^4$ is the clump-to-ambient density contrast ratio \citep{Fesen2001}.


\end{CJK*}
\end{document}